\documentclass[11pt,a4paper]{article}
\pdfoutput=1 
\usepackage{jheppub}	
\usepackage{tikz}
\usetikzlibrary{positioning,arrows}
\usetikzlibrary{decorations.pathmorphing}
\usetikzlibrary{decorations.markings}
\usetikzlibrary{snakes}
\usetikzlibrary{matrix}

\usepackage{amsmath}
\usepackage{bm}
\usepackage[utf8]{inputenc}
\usepackage[english]{babel}
\usepackage{makeidx}
\usepackage{tikz}
\tikzset{every picture/.style={line width=0.75pt}} 
\usepackage{amsfonts}
\usepackage{enumerate}
\usepackage{mathrsfs}
\usepackage{tensor}
\usepackage[autostyle]{csquotes}
\usepackage{subfig}
\usepackage{amsmath,amssymb}
\usepackage{latexsym}
\usepackage{slashed}
\usepackage{bbm}
\def\be{\begin{equation}}
\def\ee{\end{equation}}
\def\bea{\begin{eqnarray}}
\def\eea{\end{eqnarray}}
\newcommand{\nn}{\nonumber}

\newcommand{\ft}[2]{{\textstyle\frac{#1}{#2}}}

\def\apjl{\ref@jnl{ApJ}}

\def\be{\begin{equation}}
\def\ee{\end{equation}}
\def\bea{\begin{eqnarray}}
\def\eea{\end{eqnarray}}

\newcommand{\Z}{\mathbbm{Z}}

\title{On irregular states and Argyres-Douglas theories}

\author[a]{Francesco Fucito,}
\author[a]{Jose Francisco Morales,}
\author[b]{Rubik Poghossian.}
\affiliation[a]{Dipartimento di Fisica, Università di Roma ``Tor Vergata"  \& Sezione INFN Roma2, Via della ricerca
scientifica 1, 00133, Roma, Italy}
\affiliation[b]{Yerevan Physics Institute,
Alikhanian Br. 2, 0036 Yerevan, Armenia}
\abstract{Conformal theories of the Argyres-Douglas type are notoriously hard to study given that they are isolated and strongly coupled thus lacking a lagrangian description. In flat space, an exact description is provided by the Seiberg-Witten theory. Turning on a $\Omega$-background makes the geometry ``quantum" and tractable only in the weak curvature limit. In this paper we use the AGT correspondence to derive $\Omega$-exact formulae for the partition function, in the nearby of monopole points where the dynamics is described by irregular conformal blocks of the CFT. The results are checked against those obtained by the recursion relations coming from a conformal anomaly in the region where the two approaches overlap. The Nekrasov-Shatashvili limit is also discussed.  Finally, we comment on the existence of black holes in De Sitter space whose low energy dynamics is described by an  Argyres-Douglas theory.
}
 \clearpage
\begin{document}
\tikzset{
line/.style={thick, decorate, draw=black,}
 }

\maketitle


\section{Introduction}

In \cite{Argyres:1995jj},  the authors remarked that the moduli space of ${\cal N}=2$ supersymmetric gauge theories (SYM) include
points where monopoles and dyons become massless simultaneously leading to superconformal  theories (AD) isolated and strongly coupled. Being strongly coupled and involving mutually non-local massless particles, a lagrangian description is not available but the Seiberg-Witten methods devised in \cite{Seiberg:1994rs,Seiberg:1994aj} could be used to infer the behaviour of the theory. These methods were further enriched by localization \cite{Nekrasov:2002qd,Flume:2002az,Bruzzo:2002xf,Nekrasov:2003rj,Flume:2004rp} which allowed a direct computation of the non perturbative sector using field theory methods. A major role in localization is played by the extension of the theory to the so called $\Omega$-background, a curved spacetime that lifts all flat directions localizing integrals around a finite number of points  \cite{Nekrasov:2003rj,Billo:2006jm}. The results were applied in \cite{Gaiotto:2009we,Alday:2009aq} to build a correspondence (AGT) between SYM and Conformal Field Theories (CFT) in two dimensions.  The prototypical example relates
the partition function of ${\cal N}=2$ SYM with gauge group $SU(2)$ and four massless flavours to a four-point correlator in Liouville theory.

  In the framework of the AGT correspondence, AD theories correspond to degenerated limits of the correlator where three or more vertex insertions collide at a point \cite{Gaiotto:2009ma}. The partition function in this limit can be written in terms of  irregular states of the two-dimensional conformal field theory. These irregular states were made explicit in \cite{Bonelli:2011aa,Gaiotto:2012sf}
where their corresponding Verma modules were detailed and their connection with the AD theories was clarified.
These methods were further explored in \cite{Nishinaka:2012kn,Kanno:2013vi,Nishinaka:2019nuy,Kimura:2020krd,Bonelli:2021uvf,Bonelli:2022ten,Kimura:2022yua,Consoli:2022eey}.

In this paper we compute the partition function of AD theories in a generic $\Omega$-background (parameterized by $\epsilon_1,\epsilon_2$) using the formalism of irregular states.
Our motivations for this study have their roots in \cite{Grassi:2019txd,Bissi:2021rei} in which a computation of the OPE coefficients for the four space time dimensional AD SCFT was carried out. Even though these results, relying on a semi-classical approximation where the $\Omega$-background is turned off, are in good agreement with the numerical ones for the AD theories ${\cal H}_1$ and ${\cal H}_2$ \footnote{The numerical results are obtained from the conformal bootstrap method \cite{Cornagliotto:2017snu,Gimenez-Grau:2020jrx}}, in the case of the ${\cal H}_0$ theory this is not so suggesting that curvature contributions cannot be simply ignored.

At the moment the two tools available to carry out these computations are the recursion relations coming from the conformal anomaly \cite{Bershadsky:1993ta, Huang:2006si,Huang:2009md,Huang:2011qx,Huang:2013eja,Krefl:2010fm} and the use of irregular states detailed in \cite{Gaiotto:2012sf}. The former method will be the subject of a separate paper \cite{Fucito:2023txg} and provides exact formulae for the $\Omega$-deformed prepotential, order by order in the limit of small curvatures. Here we exploit the connection to  irregular states, to derive $\Omega$-exact formulae in the nearby region of monopole points where irregular states can be explicitly written.  We will show the agreement of the two methods in the region of overlap. The picture emerging from this paper and from \cite{Fucito:2023txg} is that the recursion relations coming from the modular anomaly equation can be used to explore regions both close and away from the conformal point while CFT methods, although non perturbative in the $\Omega$-background parametrization, provide an accurate description only of the region far away from the AD point, i.e. that corresponding to large expectation values of the coupling associated to the Coulomb branch operator.

We will also treat the case of the Nekrasov-Shatashvili $\Omega$-background (NS limit in which $\epsilon_1\to 0$ and $\epsilon_2\neq 0$) \cite{Nekrasov:2009rc}. This case is of interest for the study of black holes with CFT methods \cite{Aminov:2020yma,Bonelli:2021uvf,Bonelli:2022ten,Consoli:2022eey,Bianchi:2021xpr,Bianchi:2021mft}.
In particular we will show that certain black holes in De Sitter spaces admit a description in terms of AD theories.

This is the plan of the paper: in Section 2 we will discuss irregular states. in Section 3 we will compare the results obtained for the non perturbative correction of SYM from the conformal anomaly to those obtained from the irregular states. In Section 4 we will discuss the NS limit of our results. Finally in Section 5 we will draw some conclusions.

\section{Irregular conformal blocks}

\subsection{Irregular states}

Irregular conformal blocks are obtained from the conformal blocks of N-point correlators of primary fields by taking the limit where two or more primaries
collide at a point. Consider for example the collision of $n+1$ primaries with charges $\alpha_i$
and conformal dimension
\be
\Delta_{\alpha_i} =\alpha_i (Q-\alpha_i)
\ee
at the origin.
The irregular state is defined by sending all positions
$z_i \to 0$ and charges $\alpha_i \to \infty$ keeping finite the combinations
\be
c_m=\sum_{i=0}^n  \alpha_i \, z_i^m. \qquad , \qquad m=0,\ldots n
\ee
The resulting $(n+1)$-point state can be labelled by two sets of charges
\be
   {\bf c} =(c_0,c_1,\ldots c_n) \qquad , \qquad   \bm{\beta}=(\beta_0,\beta_1\ldots ,\beta_{n-1} )
\ee
and written as
\be
|I_n( {\bf c},{\bm \beta} )\rangle =\lim_{z_i\to 0} \prod_{i=0}^n  V_{\alpha_i} (z_i) |0\rangle=
\lim_{z_i\to 0}
\begin{tikzpicture}[baseline={(current bounding box.center)}, node distance=0.8cm and 0.8cm]
\coordinate[label=above:$\alpha_0$] (k0);
\coordinate[below=of k0] (s0);
\coordinate[left=of s0] (p0);
\coordinate[right=1cm of s0] (s1);
\coordinate[above=of s1,label=above:$\alpha_{1}$] (k1);
\coordinate[right=of s1] (p2);
\coordinate[right=1cm of s1] (s2);
\coordinate[right=1cm of s2] (s2bis);
\coordinate[right=1cm of s2bis] (s2tris);
\coordinate[above=of s2tris] (k2tris);
\coordinate[above=of s2bis,label=above:$\alpha_{n-2}$] (k2bis);
\coordinate[above=of s2tris,label=above:$\alpha_{n-1}$] (k2tris);
\coordinate[right=of s2tris] (p3);
\draw[line] (k0) -- (s0);
\draw[line,dashed] (s0) --node[label={[xshift=0.2cm, yshift=0.1cm]left:\scriptsize{$ z_0$}}] {}  (k0);
\draw[line] (s0) -- (p0);
\draw[line] (k1) -- node[label={[xshift=0.2cm, yshift=0.1cm]left:\scriptsize{$ z_{1}$}}] {} (s1);
\draw[line] (s1) -- (p2);
\draw[line] (s0) -- node[label={[yshift=0.2cm]below:$\beta_0$}] {} (p0);
\draw[line] (s0) -- node[label={[yshift=0.2cm]below:}] {} (s1);
\draw[line] (s1) -- node[label={[yshift=0.2cm]below:}] {} (s2);
\draw[line,dashed]  (s2) -- (s2bis);
\draw[line] (s2bis) -- node[label={[yshift=0.2cm]below:}] {} (s2tris);

\draw[line] (k2bis) -- node[label={[xshift=0.2cm, yshift=0.1cm]left:\scriptsize{$ z_{n-2}$}}] {} (s2bis);
\draw[line] (k2tris) -- node[label={[xshift=0.2cm, yshift=0.1cm]left:\scriptsize{$ z_{n-1}$}}] {} (s2tris);

\draw[line] (s2tris) -- node[label={[yshift=0.2cm]below:$\alpha_n$}] {}  (p3);
\label{fusb2}
\end{tikzpicture}
\ee
The set $\{\beta_1\ldots ,\beta_{n-1}\}$ parameterizes the momenta flowing in the intermediate states in \eqref{fusb2} and are related to the vacuum expectation values of the adjoint scalars in the associated quiver gauge theory.
Alternatively the $ |I_n\rangle $  state can be defined by the eigenvalue equations \cite{Gaiotto:2012sf}
\bea
 L_k |I_n( {\bf c},{\bm \beta} )\rangle  &=& {\cal L}^{(n)}_k|I_n( {\bf c},{\bm \beta} )\rangle \qquad ,\qquad k=0,\ldots n \nn\\
  L_k |I_n( {\bf c},{\bm \beta} )\rangle  &=& \Lambda^{(n)}_k|I_n( {\bf c},{\bm \beta} )\rangle \qquad ,\qquad k=n{+}1,\ldots 2n\nn\\
  L_k |I_n( {\bf c},{\bm \beta} )\rangle  &=& 0  \qquad ,\qquad\qquad\qquad ~~~~k> 2n
  \label{lneq}
\eea
where the $\{L_k\}_{k\in\Z}$ are the generators of the Virasoro algebra and
\bea
\Lambda^{(n)}_k & =& \left[ (k+1)Q-2c_0\right] c_k -\sum_{\ell=1}^{k-1} c_\ell \, c_{k-\ell} \qquad , \qquad k=n+1,\ldots 2n \nn\\
{\cal L}^{(n)}_k   &=&   \Lambda_k +\sum_{\ell=k+1}^{n-1} (l- k)  c_l {\partial \over \partial c_{l-k} }  \qquad , \qquad \qquad  k=1,\ldots n \nn\\
{\cal L}^{(n)}_0   &=&  (Q-c_0) c_0 +\sum_{k=1}^n k  c_k {\partial \over \partial c_k} \label{lambdak}
\eea
  The state  $|I_n({\bf c},{\bf \beta} )\rangle $ can be obtained recursively starting from the lowest
ones following the ansatz
\be
|I_n( {\bf c},{\bm \beta} )\rangle  =
f({\bf c}, \beta_{n-1} )  \sum_{k=0}^\infty c_n^k   |I_{n-1}( \tilde {\bf c},\tilde {\bm\beta} ),k \rangle
\ee
with
\be
    \tilde{\bf c} =(\beta_{n-1}  ,c_1,\ldots c_{n-1} ) \qquad , \qquad   \tilde{\beta}=(\beta_0,\beta_1\ldots \beta_{n-2})
\ee
$  |I_{n-1}( \tilde {\bf c},\tilde {\bm\beta} ),k \rangle $ is a generalized descendant of level $k$ of  $ |I_{n-1}( \tilde {\bf c},\tilde {\bm\beta} )\rangle $ obtained by acting with Virasoro generators and derivatives with respect to $c_1,\ldots c_{n-1}$ on the descendant of level $k$. The precise form of these descendants can be determined order by order solving the eigenvalue equations (\ref{lneq}).

\subsection{Irregular conformal blocks}

Given a pair of irregular states, we define the partition function
\be
Z_{mn}= \langle I_m( {\bf c}',{\bm \beta}' )  |I_n( {\bf c},{\bm \beta} )\rangle. \qquad  m,n\geq -1
\label{irregularpartition}\ee
that represents the irregular limit of the $(n+m+2)$-point conformal block where primaries collide at two point $z=\infty$ and $z=0$ in two groups of $(m+1)$ and $(n+1)$ primaries. We take always $n\geq m$.
Here and in the following, the standard vacuum and the primary state of conformal dimension $\Delta_{c_0}$ are denoted by
\be
 |I_0( c_0 )\rangle=| \Delta_{c_0} \rangle \qquad, \qquad  |I_{-1} \rangle= | 0 \rangle
\ee
Using (\ref{lneq}), (\ref{lambdak})  we can introduce the meromorphic function
\be
\phi_2(z)   =- { \langle  I_m | T(z) | I_n \rangle  \over \langle I_m  | I_n\rangle } =  \sum_{k=-2n-2}^{2m-2}  v_k \, z^k
\ee
with
\bea
v_{k}=
\left\{
\begin{array}{lll}
   \Lambda^{(n)}_{-k}  &~~~~  & k= -(n+1),\ldots -2n\\
  {\cal L}^{(n)}_{-k}  \ln Z_{mn} &&  k= -1,-2,\ldots -n\\
 {\cal L}^{(m)}_k \ln Z_{mn} &  &k= 1,2,\ldots m \\
 \Lambda^{(m)}_{k}  & & k= (m+1),\ldots 2m
\end{array}
\right.
\label{vks}
\eea
The parameters $v_k$ with $-n\leq k \leq m$ and $k \neq 0$ will be identified with the Coulomb branch parameters of an AD theory with $\Omega$-deformed partition function $Z_{mn}$. According to (\ref{vks}) they are always given by derivatives of the AD prepotential
\be
{\cal F}_{mn}=p \log Z_{mn}
\ee
with $p=\epsilon_1 \epsilon_2$ and $\epsilon_1$, $\epsilon_2$ the parameters specifying the $\Omega$-background. These relations generalize the quantum Matone equation \cite{Matone:1995rx,Flume:2004rp} to the AD case.
 The SW differential will be related to $\phi_2(z)$ via
\be
\label{SW_diff_vs_phi_2}
\lambda=\lim_{\epsilon_{1,2} \to 0} \sqrt{p\phi_2(z) } dz
\ee
 Without loss of
 generality we can reabsorb $v_{-2n-2}$ into a rescaling of $z$, and  assign conformal dimension one to the SW differential.
 This leads to the dimensions
\bea
 [z]=-{1\over n}  \qquad , \qquad  [ v_k ] &=& 2+ { k \over n}   \qquad, \qquad -2n\leq k \leq 2m
 \eea
 From the six-dimensional point of view, the choices $m=-1$, $m=0$ and $m>0$ arise from compactifications on an sphere with a single irregular puncture, one regular and one irregular puncture and two irregular punctures. We will focus on the first two cases corresponding to AD theories of type $(A_1,A_{2n-3})$ and $(A_1,D_{2n} )$ respectively.

 \subsection*{ $m=-1$: the  $(A_1,A_{2n-3})$ series}

The type AA series is obtained by taking $m=-1$ leading to
\bea
 \phi_2  &=& -{ \langle  0 | T(z) | I_n \rangle  \over \langle 0  | I_n\rangle } = {\Lambda_{2n}^{(n)}\over z^{2n+2} }+  \ldots
    + {\Lambda_{n+1}^{(n)}\over z^{n+3} }+{v_{-n}\over z^{n+2} }+ \ldots + {v_{-2} \over z^4}
    \eea
    where we use that $v_{-1}=v_0=0$ since $L_{1} |0 \rangle =L_{0} |0 \rangle=0$.
   We can always set  $ \Lambda_{2n}^{(n)}=1$ and $\Lambda_{2n-1}=0$. The remaining parameters have dimensions
 \bea
 n=3:~(A_1,A_3) &&  [v_{-2}]=\ft43 \quad, \quad [v_{-3} ]=1 \quad, \quad [\Lambda^{(3)}_4]=\ft23 \\
 n=4:~(A_1,A_5) &&  [v_{-2}]=\ft32 \quad, \quad  [v_{-3} ]=\ft54  \quad, \quad [v_{-4} ]=1 \quad, \quad [\Lambda^{(4)}_5]=\ft34  \quad, \quad [\Lambda^{(4)}_6]=\ft12  \nn
 \eea
 Analogous formulae hold for the cases with $n>4$.

 \subsection*{ $m=0$: the  $(A_1,D_{2n})$ series}

The type AD series is obtained by taking $m=0$ leading to
\bea
 \phi_2 &=& - { \langle  \Delta_\alpha | T(z) | I_n \rangle  \over \langle \Delta_\alpha  | I_n\rangle }   = {\Lambda_{2n}^{(n)}\over z^{2n+2} }+  \ldots
    + {\Lambda_{n+1}^{(n)}\over z^{n+3} }+{v_{-n}\over z^{n+2} }+ \ldots + {v_{0} \over z^2}
\eea
 where we use that $v_1=0$ since $L_{1} | \Delta_\alpha \rangle =0$.
 We can always set   $ \Lambda_{2n}^{(n)}=1$. The remaining parameters have dimensions
 \bea
 n=2:~(A_1,D_4) &&  [v_{0} ]=2  \quad, \quad  [v_{-1}]=\ft32  \quad, \quad [v_{-2} ]=1  \quad, \quad [\Lambda^{(2)}_3]=\ft12  \\
 n=3:~(A_1,D_6) &&  [v_0]=2 \quad, \quad  [v_{-1}]=\ft53  \quad, \quad [ v_{-2} ]=\ft43  \quad, \quad [ v_{-3} ]=1 \quad, \quad  [\Lambda_4]=\ft23  \quad, \quad [\Lambda_5]=\ft13  \nn
 \eea
 Analogous formulae hold for the cases with $n>3$.

   \subsection*{ $m=1$: A  new series}

 Next we consider $m=1$. One finds
 \bea
  \phi_2  &=& - { \langle  \Delta_\alpha | T(z) | I_n \rangle  \over \langle \Delta_\alpha  | I_n\rangle }   = {\Lambda_{2n}^{(n)}\over z^{2n+2} }+  \ldots
    + {\Lambda_{n+1}^{(n)}\over z^{n+3} }+{v_{-n}\over z^{n+2} }+ \ldots +  {v_{1} \over z}  +\tilde \Lambda^{(1)}_2
\eea
 We can always set   $ \Lambda_{2n}^{(n)}=1$. The remaining parameters have dimensions
 \bea
 n=1: &&  [\tilde \Lambda^{(1)}_{2} ]=4  \quad, \quad  [v_{1}]=3  \quad, \quad [v_{0} ]=2
 \quad, \quad [v_{-1} ]=1   \\
 n=2: &&  [\tilde \Lambda^{(1)}_{2} ]=3  \quad, \quad  [v_{1}]=\ft52  \quad, \quad [v_{0} ]=2
 \quad, \quad [v_{-1} ]=\ft32  \quad, \quad [v_{-2} ]=1  \quad, \quad [\Lambda^{(2)}_{3} ]=\ft12  \nn
 \eea
 Analogous formulae hold for the cases with $n>2$. The $n=1$ case corresponds to $SU(2)$ gauge theory with two flavours  and the four parameters are related to the Coulomb branch parameters, the renormalization group invariant scale and two masses. The case $n=2$  corresponds to a partial $SU(2)\in SU(3)$ gauging of the flavour symmetry of the ${\cal H}_2$ theory\footnote{We thank S. Giacomelli and R. Savelli for suggesting us this dictionary. }.

\subsection{ ${\cal H}_1$ AD theory}

The partition function of the ${\cal H}_1$ AD theory is given by
\be
Z_{{\cal H}_1} =\langle  0  |I_3({\bf c}, {\bm \beta})\rangle \label{zh1}
\ee
  Then one finds
\bea
\phi_2(z) &=&- { \langle 0 | T(z) | I_3 \rangle  \over \langle 0 | I_3\rangle } =
{ 2v \over z^4}  {+}{ 2( c_1 c_2 +c_3 (c_0-2Q) ) \over z^5}
{+}{c_2^2{+}2 c_1 c_3 \over z^6}
{+}{ 2 c_2 c_3 \over z^7}{+}{c_3^2\over z^8} \label{phi2h1}
\eea
with
\be
v=-\frac{c_3}{2} \partial_{c_1} \log Z_{{\cal H}_1}  {+}c_2 \left(c_0- {3Q \over 2}\right)
+{c_1^2 \over 2} \label{vh1}
\ee
In the limit $c_3\ll 1$, one finds
\be
 |I_3({\bf c}, {\bm \beta})\rangle  =c_2^{\rho_2 }c_3^{\rho_3 } e^{(c_0-\beta_2) \left({2 c_1 c_2 \over c_3} -{c_2^3 \over 3 c_3^2}-{c_1^2 \over c_2} \right)  }  \sum_{k=0}^\infty c_3^k  |I_2(\tilde {\bf c}, \tilde {\bm \beta}),3k \rangle
 \label{i3c}
\ee
with
\bea
\rho_2 &=& \ft12(\beta_2-c_0)(5\beta_2-c_0-5Q) \qquad, \qquad \rho_3 = 2(\beta_2-c_0)(Q-\beta_2) \nn\\
\tilde {\bf c} &=&(\beta_2,c_1,c_2) \qquad , \qquad \tilde {\bm \beta}=(\beta_0,\beta_1)
\eea
 Plugging (\ref{i3c}) into (\ref{zh1}) one finds $Z_{{\cal H}_1}=Z_{{\cal H}_1,\rm tree}Z_{{\cal H}_1,\rm inst}$ with
 \bea
 \label{Zh1tree}
 Z_{ {\cal H}_1 \rm tree} &=& c_2^{\rho_2-\frac{1}{2} \beta _2 \left(Q-\beta _2\right)}c_3^{\rho_3} e^{(c_0-\beta_2) \left({2 c_1 c_2 \over c_3} {-}{c_2^3 \over 3 c_3^2}{-}{c_1^2 \over c_2} \right)+{c_1^2\over c_2} (\beta_2-Q)  }\\
 Z_{ {\cal H}_1 \rm inst} &=&
  1{-}c_3 \left(  {2 c_1^3\over 3 c_2^3} (Q{+}c_0{-}2\beta_2) {+}{c_1\over c_2^2} ( c_0^2 {+} 6 \beta_2(\beta_2-c_0) {+}Q (5c_0{-} 6\beta_2) )  \right){+}\ldots
   \eea
   For the sake of simplicity, the terms with higher powers of $c_3$, which are needed for the comparison with the gauge theory results, are omitted here and can be found in the appendix.
 We can check these results against a SW analysis. To this aim we introduce the variables
  \be
 \label{map_CFT_H1}
 Q=\frac{s}{\sqrt{p}};~ c_0=\frac{\hat{c}_0+2s}{\sqrt{p}};~ \beta_2=\frac{3s+2a}{2\sqrt{p}}; ~v ={\hat{v} \over p} ; ~\phi_2=\frac{\hat\phi_2}{p} ~  c_i=\frac{\hat{c}_i}{\sqrt{p}};~i=1,2,3
 \ee
 where $s=\epsilon_1+\epsilon_2$ and  $p=\epsilon_1\epsilon_2$. We then have
 \bea
 \hat\phi_2(z) &=&
 { 2\hat{v} \over z^4}  {+}{ 2( \hat{c}_1 \hat{c}_2 +\hat{c}_3 \hat{c}_0 ) \over z^5}
 {+}{\hat{c}_2^2{+}2 \hat{c}_1 \hat{c}_3 \over z^6}
 {+}{ 2 \hat{c}_2 \hat{c}_3 \over z^7}{+}{\hat{c}_3^2\over z^8} \label{phi2hath1}
 \eea
 and
 \be
 a ={1\over 2\pi i} \oint_{z=0} \sqrt{\hat\phi_2} dz  =
 \frac{2 \hat{v}-\hat{c}_1^2}{2 \hat{c}_2}-\frac{\hat{c}_1 \hat{c}_3
 	\left(\hat{c}_1^2+\hat{c}_0 \hat{c}_2-2 \hat{v}\right)}{\hat{c}_2^3}+\ldots
 \ee
 Inverting for $\hat{v}$ one finds
 \be
 \hat{v}=a \hat{c}_2+\frac{\hat{c}_1^2}{2}
 +\frac{\hat{c}_3 \hat{c}_1 \left(\hat{c}_0-2a\right)}{\hat{c}_2}+\ldots
 \ee
 This matches the result for $v$ given by plugging (\ref{Zh1tree}) into (\ref{vh1})
 \be
 v=\beta _2 c_2+\frac{c_1^2}{2}-\frac{3c_2Q}{2}+\frac{c_3 c_1 \left(c_0+Q-2 \beta _2\right)}{c_2}+\ldots
 \ee
 taking into account (\ref{map_CFT_H1}). In the following sections we will show that this agreement holds also in presence of $\epsilon$-corrections.

\subsection{ ${\cal H}_2$ AD theory}

The partition function of the ${\cal H}_2$ AD theory is given by
\be
Z_{{\cal H}_2} =\langle \beta_0  |I_2({\bf c}, {\bm \beta})\rangle  \label{zh2}
\ee
We define
\bea
\phi_2(z)&=& -{ \langle \Delta_{\beta_0} | T(z) | I_2\rangle  \over \langle \Delta_{\beta_0}  | I_2\rangle } =
-{\Delta_{\beta_0}\over z^2}  {+}{2 v \over z^3}
{+}{c_1^2{+}c_2(2 c_0{-}3Q)\over z^4} {+}{2c_1 c_2 \over z^5}
{+}{c_2^2\over z^6} \label{phi2h2}
\eea
where the right hand side is computed using (\ref{lneq}) and\footnote{This is the analog
of the relation \cite{Flume:2004rp} for the case of the AD-theory .}
\be
v={-}{c_2\over 2}\, \partial_{c_1} \ln Z_{{\cal H}_2} {+}c_1 (c_0{-} Q)  \label{vh2}
\ee
The partition function can be computed in the limit $c_2 \ll 1$ where the irregular state  $I_2$ can be written in terms of $I_1$  (see appendix for details)
\be
|I_2({\bf c}, {\bm \beta})\rangle  =c_1^{\nu_1 }c_2^{\nu_2 } e^{(c_0-\beta_1){c_1^2\over c_2}  } \left[  1{+} c_2\left( {\nu_3\over c_1^2} {+}{(2 c_0 {-}3 \beta_1)\over c_1} \partial_{c_1}
{+}{L_{-1}\over 2 c_1}  \right) {+}\ldots \right]  |I_1(\beta_1,\tilde {\bf c}, \tilde {\bm \beta})\rangle \label{i2c}
\ee
with
\bea
\nu_1 &=& 2(c_0 -\beta_1)(Q-\beta_1) \qquad, \qquad \nu_2 = \ft12(\beta_1-c_0)(3Q-3 \beta_1-c_0) \nn\\
\nu_3 &=& \ft12 (c_0 + 3 Q - 3 \beta_1) (c_0 - \beta_1) (Q - \beta_1)\nn\\
\tilde {\bf c} &=&(\beta_1,c_1) \qquad , \qquad \tilde {\bm \beta}=\beta_0
\eea
Plugging (\ref{i2c}) into (\ref{zh2}) one finds
$Z_{{\cal H}_2}=Z_{{\cal H}_2,\rm tree}Z_{{\cal H}_2,\rm inst}$ with
\bea
\label{Zh2tree}
Z_{ {\cal H}_2 \rm tree} &=&  c_1^{\nu_1 +\Delta_{\beta_0}-\Delta_{\beta_1} }c_2^{\nu_2 } \, e^{(c_0-\beta_1){c_1^2\over c_2}  }\\
Z_{ {\cal H}_2 \rm inst} &=&
1{+} {c_2\over 2c_1^2} \left[2 \nu_3  {+}(2 c_0{-}3 \beta_1) (\beta_1-\beta_0)(\beta_1+\beta_0-Q)\right]
{+}\ldots
\eea
For the sake of simplicity, the terms with higher powers of $c_2$, which are needed for the comparison with the gauge theory results, are omitted here and can be found in the appendix.
We can check these results against a SW analysis. To this aim, we should go from
the CFT variables to
\be
\label{map_CFT_H2}
Q=\frac{s}{\sqrt{p}};~
c_0=\frac{2\hat{c}_0+3s}{2\sqrt{p}}; ~\Delta_{\beta_0}=-\frac{M^2}{p};~ \beta_1=\frac{s+a}{\sqrt{p}}
;~\quad v=\frac{\hat{v}}{p}  ;~\phi_2=\frac{\hat\phi_2}{p};~ c_i=\frac{\hat{c}_i}{\sqrt{p}}~i=1,2\,;
\ee
Then
\bea
\hat\phi_2(z)&=&
{M^2\over z^2}  {+}{2\hat{v} \over z^3}
{+}{\hat{c}_1^2{+}2\hat{c}_0\hat{c}_2\over z^4} {+}{2\hat{c}_1 \hat{c}_2 \over z^5}
{+}{\hat{c}_2^2\over z^6} \label{phi2h2hat}
\eea
and
\be
a ={1\over 2\pi i} \oint_{z=0} \sqrt{\hat\phi_2(z) } dz
=-\frac{\hat{c}_2 \left(M^2\hat{c}_1^2+2 \hat{c}_0 \hat{c}_1 \hat{v}
	-3 \hat{v}^2\right)}{2\hat{c}_1^4}+\frac{\hat{v}}{\hat{c}_1}+\ldots
\ee
Inverting for $\hat{v}$ one finds
\be
\hat{v}=a \hat{c}_1-\frac{\hat{c}_2 \left(3 a^2-2 a \hat{c}_0-M^2\right)}{2 \hat{c}_1}+\ldots
\ee
This matches the result for $v$ given by plugging (\ref{Zh2tree}) into (\ref{vh2})
\be
v=(\beta _1-Q) c_1+\frac{c_2 \left(3\Delta_{\beta_1}-\Delta_{\beta_0}+2c_0(\beta _1-Q )\right)}{2c_1}+\ldots
\ee
in view of the identification rules (\ref{map_CFT_H2}).
In the next sections we will show this match to hold also in presence of $\epsilon$-corrections.
\section{ The holomorphic anomaly recursion }

In this section we first derive formulae which are exact in the parameter of the instanton expansion in winding number for  the leading $\epsilon$-corrections of the prepotential. This provides a complementary picture with respect to that obtained in the previous sections providing $\epsilon$-exact formulae  for the partition function in the neighborhood of a degenerating point (a zero of the discriminant of the SW curve) where vertex insertions collide to produce irregular states.

We write
\bea
{\cal F}=\epsilon_1\epsilon_2 \log Z=\sum_{g,h=0}^\infty \left(\epsilon_1+\epsilon_2\right)^{2n}
\left(\epsilon_1\epsilon_2\right)^{g}F^{(n,g)} =\sum_{g=0} {\cal F}_g( b) \left(\epsilon_1\epsilon_2\right)^{g}
\eea
with
\be
b=\sqrt{\epsilon_1\over \epsilon_2}
\ee

As in the previous sections we parameterize  the $\Omega$-background with the variables
\be
s=\epsilon_1+\epsilon_2 \qquad, \qquad p=\epsilon_1 \epsilon_2
\ee

\subsection{The ${\cal F}_0$ prepotential}

The ${\cal F}_0$ term describes the SW prepotential and it is $b$-independent. It can be
encoded in the periods $\omega_i =\oint_{\gamma_i} dz/(i \pi y)$. The $\gamma_i$ are a basis of non trivial cycles encircling the end points of the cuts of the elliptic curve
\be
y^2 =4 z^3 - g_2(u) z- g_3(u)
\ee
with discriminant
\be
\Delta (u) = 16 (g_2(u)^3-27 g_3(u)^2)
\ee
We denote by $\tau= {\omega_2\over \omega_1}$, the ratio of the periods and parameterize it by the variable
$q=e^{\pi {\rm i} \tau}$.  The explicit dependence $u(q)$ and $\omega_1(u,q)$ can be obtained by solving the relations, coming from the elliptic geometry \footnote {The Eisenstein series are given by
$E_k(q)=1+{2\over \zeta (1-k)}\sum _{n=1}^{\infty} \frac{n^{k-1} q^{2 n}}{1-q^{2 n}}$\,, $k=2,4,6,\cdots $\,.}
\be
{27 g_3(u)^2\over g_2(u)^3} = { E_6(q)^2\over E_4(q)^3} \qquad , \qquad \omega_1(q,u)^2= \frac{2g_2(u)E_6(q)} {9 g_3(u)E_4(q)} \label{w1ee}
\ee
 In particular from the first equation one finds
 \be
D_\tau u\equiv q \partial_q u  = {2\left(  E_6^2-E_4^3\right) \over E_6 E_4 \left( 2 { g_2'(u) \over g_2(u) } -
 3 { g_3'(u)  \over g_3(u) } \right)  }
\ee
 where we have used (the last two of) the well known differentiation rules
 \be
 D_\tau E_2 = \ft{1}{6} (E_2^2-E_4) \quad  ,
 \quad D_\tau E_4 = \ft{2}{3} (E_2 \, E_4-E_6)  \quad , \quad
 D_\tau E_6=  E_2 \, E_6-E_4^2\,.
 \ee
 Finally, in a standard way we introduce the variable $a$ and the SW prepotential ${\cal F}(a)$ related to the variables $u$ and $q$ via
\be
 \log q =-\frac{{\cal F}''(a)}{2} \qquad , \qquad    \omega_1(q,u)={da\over du}
\ee

 \subsection{ ${\cal F}_g$-terms}

 Higher derivative terms can be computed recursively \cite{Huang:2009md,Huang:2011qx,Huang:2013eja,Billo:2013fi,Billo:2013jba,Ashok:2015cba,Billo:2015pjb,Billo:2015jyt,Krefl:2010fm,Codesido:2017dns}
 \be\label{HAE}
 \partial_{E_2} {\cal F}_{g}  =\ft{1}{24} \left[    \partial_a^2 {\cal F}_{g-1} +\sum_{g'=1}^{g-1} \partial_a {\cal F}_{g'} \partial_a {\cal F}_{g-g'}  \right]
\ee
 starting from the lowest one
\be
 {\cal F}_1 (u,b,q)=-{1 \over 2}  \log \omega_1(q,u) +\ft{ b^2+b^{-2} }{24}\log \Delta (u) \label{f1h1}
 \ee
 Following \cite{Huang:2006si,Huang:2009md,Huang:2011qx,Huang:2013eja} it is convenient to introduce the variables
 \be
S(u)={1\over \omega_1(q,u)^2 }=\frac{g_3(u)E_4(q)}{g_2(u)E_6(q)}  \qquad ,\qquad  X(u) = S(u)  E_2(q)
\ee
and their derivatives
\bea
p_1 &=& {d\over du} \ln S(u) =\frac{9 X \left(2 g_2 g_3'-3 g_3 g_2'\right)+g_2^2 g_2'-18 g_3 g_3'}{2 \left(g_2^3-27
   g_3^2\right)} \\
p_2 &=&{dX(u)\over du}  =\frac{27 X^2 \left(2 g_2 g_3'-3 g_3 g_2'\right)+6 X \left(g_2^2 g_2'-18 g_3
   g_3'\right)+g_2 \left(2 g_2 g_3'-3 g_3 g_2'\right)}{12 \left(g_2^3-27 g_3^2\right)}\nn
\eea
with
\be
{d\over du} =\partial_u + {1\over D_\tau u} D_\tau + p_2 \partial_X
\ee
In terms of these variables one can write the $a$-derivative of $u$ as
\bea
\left(\frac{du}{da}\right)^2 &=& {1 \over  \omega_1^{2} }
=\frac{9S}{2}\qquad , \qquad \frac{d^2u}{da^2} =\frac{1}{2}\frac{d}{d u}\, \omega_1^{-2}
=\frac{9S\,p_1}{4}  \label{dau}
\eea
 and bring the holomorphic equation (\ref{HAE}) to the form
\bea\label{HAE2}
 \partial_{X} {\cal F}_{g}
 &=&  \frac{3}{16} \left[  {d^2 {\cal F}_{g-1} \over du^2}  + {p_1\over 2}   {d {\cal F}_{g-1} \over du}
 +  \sum_{g'=1}^{g-1}  {d  {\cal F}_{g'}  \over du} { d {\cal F}_{g-g'}  \over du}  \right]\
\eea
 The $ {\cal F}_{g}  $-term is obtained by integrating over $X$ the right hand side of (\ref{HAE}) and fixing the $E_2$-independent part $h_g(q,u)$ by imposing, near each zeros $u_*$ of the discriminant, the gap conditions (for $g>1$)
 \bea\label{GeneralBernoulli}
  {\cal F}_g ( b,u,q)   & \underset{ u\to  u^*} {\approx}  &  {(2g-3)! \sum_{k=0}^{g} \hat B_{2k} \hat B_{2g-2k}\frac{\epsilon_1^{2g-2k}\epsilon_2^{2k} }{a^{2g-2}}}  + O( a^0)
  \eea
  with
  \be
  \hat{B}_m=\left( 2^{1-m}-1\right) {B_m\over m!}
  \ee
  and $B_m$ the Bernoulli numbers.
   In the next sections we apply the recursion algorithm to derive $q$-exact formulae for  ${\cal F}_{2}$ in the cases
  of  the ${\cal H}_1$ and  ${\cal H}_2$ theories and show the equivalence with the results obtained from the irregular states.

\subsection{${\cal H}_1$ theory}
 In this section we  derive $q$-exact formulae for the first few ${\cal F}_g$-terms using  the holomorphic recursive algorithm. The results
 will be checked against those obtained in the previous section using irregular states.
   The dynamics of the ${\cal H}_1$ theory is described by the elliptic curve
 \bea
w^2  =\sum_{i=0}^4d_iz^{4-i} =-\left(2u\, z^4+ \mu \,z^3+c z^2+1\right)
\eea
or equivalently in Weierstrass form
 \be
 y^2 =4 x^3 -g_2(u) x-g_3(u)
 \ee
 with
 \bea
g_2&=& \frac{4 d_2^2}{3}-4 d_1 d_3+16 d_0 d_4=32 u +  \ft43 c^2\nn\\
 g_3 &=&
-\frac{1}{27} 8 d_2^3+\frac{4}{3} d_1 d_3 d_2
+\frac{32}{3} d_0 d_4 d_2-4 d_0 d_3^2-4 d_1^2 d_4
=4\mu^2 -{64 c u\over 3} +{8c^3\over 27}   \label{g2g3h1}
\eea
 For simplicity we take $\mu=0$. For this choice, the discriminant reduces to
 \be
 \Delta=8192 u(c^2-8 u)^2
 \ee
    Plugging (\ref{g2g3h1}) into (\ref{w1ee}), (\ref{f1h1}) and (\ref{HAE}), one finds
  \bea
  &&-\ft12{\partial^2 {\cal F}_0 \over \partial a^2} = \log q \\
 && p {\cal F}_1 ={p \over 4}  \log \left[ \frac{ X(u)}{E_2(q) } \right] +\ft{ s^2-2p }{24 }\log  \left[
8192 u(c^2-8 u)^2 \right] \nn\\
&& p^2 {\cal F}_2 = \frac{1}{24 u^2 \left(c^2-8 u\right)^2}  \left[ X(u)^3 \frac{135 c^2 p^2}{512}
{+}X(u)^2 \left(p^2 \left(\frac{99 c^3}{256}-\frac{189 c u}{32}\right)
+p s^2 \left(\frac{27 c u}{16}-\frac{9 c^3}{128}\right)\right) \right. \nn\\
 &&
 +X(u) \left(p^2 \left(\frac{63 u^2}{2}-\frac{9 c^2 u}{4}+\frac{25 c^4}{128}
 \right)-p s^2 \left(36 u^2-3 c^2 u
 +\frac{3 c^4}{16}\right)+s^4 \left(\frac{9 u^2}{2}-\frac{3 c^2 u}{8}
 +\frac{c^4}{128}\right)\right)
  \nn\\
   && \left.
  -p^2 \left(7 c u^2+\frac{5 c^3 u}{12}-\frac{299 c^5}{8640}\right)
  +p s^2 \left(14 c u^2+c^3 u-\frac{103 c^5}{1440}\right)-s^4 \left(5 c u^2+\frac{5 c^3 u}{12}
  -\frac{79 c^5}{2880}\right) \right] \nn
  \eea
 with
 \be
  X(u)=  \frac{2 E_4 E_2 \left(c^3-72 c u\right)}{9 E_6 \left(c^2+24 u\right)}
  \ee
    The $X$-independent part has been fixed  imposing the gap conditions
  \bea
  p^2 {\cal F}_2 & \underset{u\to c^2/8}  \approx &  {2 \over a^2} \times \left( \frac{p^2}{240 }
  -\frac{7 s^2 p }{1440}  +   \frac{7 s^4}{5760  } \right) +O(a^0) \nn\\
   p^2 {\cal F}_2 & \underset{u\to 0} \approx &  {1 \over a^2} \times
   \left( \frac{p^2}{240 }-\frac{7 s^2 p }{1440}  +   \frac{7 s^4}{5760  } \right) +O(a^0)
  \eea
  near the two degenerated points. The extra factor of two comes from the fact that $u= c^2/8$ is a second order zero of the discriminant.

  \subsection*{Comparison against CFT}

     To compare against the CFT results one starts from the quadratic differential $ \hat\phi_2 dz^2$ defined by  (\ref{phi2hath1}) and
    replaces
 \be
 z \to \frac{2 \hat{c}_3 z}{2 \hat{c}_3^{2/3}-\hat{c}_2 z}
 \ee
 to set to zero the $z^{-7}$ term and to 1 the coefficient of $z^{-8}$. We set
 \be
 \hat{c}_0 = {\mu\over 2} \quad, \quad \hat{c}_1 = 0  \quad, \quad
 \hat{c}_2 = -i \sqrt{2c}  \quad, \quad \hat{c}_3 = 1  \quad, \quad
 \hat{v} = u-\frac{c^2}{8}-\frac{ i  \mu\sqrt{c}  }{2 \sqrt{2}} +\ldots
 \label{dicth1}
 \ee
 where lower dots stand for corrections in $s$ and $p$.
 \eqref{dicth1} brings the quadratic differential to the standard ${\cal H}_1$ form
 \be
\hat\phi_2= {2u\over z^4} +{\mu \over z^5}+ {c\over z^6}+ {1\over z^8}
\ee
Setting the dimension $[\phi_2 dz^2]=2$ one finds the expected dimensions
 \be
[c]=\ft23 \quad, \quad [u]=\ft43 \quad\quad [\mu]=1 \quad , \quad [z] =-\ft13
 \ee
for the Coulomb branch parameter $u$, the coupling  $c$ and the mass $\mu$ of the ${\cal H}_1$-theory.
Applying the map (\ref{map_CFT_H1}) to (\ref{prepotentiah1cft}) and specifying the parameters as in
(\ref{dicth1}) one finds for the instanton prepotential
 \bea
{\cal F}^{CFT}_{{\rm inst},{\cal H}_1} &=& p \log Z^{CFT}_{{\rm inst},{\cal H}_1}
= \frac{125 a^4}{32 c^3}-\frac{17i  a^3}{6 \sqrt{2} c^{3/2}}+s^2
   \left(-\frac{19i a}{24 \sqrt{2} c^{3/2}}+\frac{153 a^2}{64 c^3} \right) \nn\\
   && +p \left(\frac{i a}{12 \sqrt{2}c^{3/2}}-\frac{15 a^2}{32 c^3}\right)
   +\frac{131 s^4}{1536 c^3}-s^2 p\,\frac{17}{384 c^3}+\ldots \label{fh1inst}
 \eea
We can compare these results against those obtained from the holomorphic
anomaly equations after taking $u$ and $c$ large and
  $u\approx c^2/8$. In this limit, the discriminant vanishes and $q\approx 0$. Expanding (\ref{w1ee}) around $q\approx 0$ and $u\approx c^2/8$ one finds
  \bea
q(a) &=& \frac{a^2}{32 c^3} \left[ 1{-}\frac{17 a}{\sqrt{2} c^{3/2}}+\frac{953 a^2}{8 c^3}-\frac{25239 a^3}{16 \sqrt{2} c^{9/2}}
+\frac{2616331 a^4}{256 c^6}
{+}\ldots \right] \nn\\
u(a)& =&   \frac{c^2}{4}+2 \sqrt{2c} \, a \left[ 1{+} \frac{3 a}{2 \sqrt{2} c^{3/2}}{-}\frac{17 a^2}{8 c^3}{+}\frac{375 a^3}{32 \sqrt{2} c^{9/2}}
{-}\frac{10689 a^4}{256 c^6} {+}\ldots \label{quh1}
\right]
  \eea
Using the expansions (\ref{quh1}) one finds
  \bea
  {\cal F}_0&=& -{a^2\over 2}  \log{-a^2 \over 32 c^3 }+\frac{3 a^2}{2}
  -\frac{17i a^3}{6 \sqrt{2} c^{3/2}} +\frac{125 a^4}{32 c^3} +\ldots \\
 p{\cal F}_1&=& p \left({-}\frac{\log 32a}{6} {+}\frac{ia}{12 \sqrt{2} c^{3/2}}
 {-}\frac{15 a^2}{32 c^3}{+}\ldots\right){+}s^2 \left(\frac{1}{24} \log {a^2\over 8192 c^3}
 {+}\frac{153 a^2}{64c^3}{-}\frac{19i a}{24 \sqrt{2} c^{3/2}}{+}\ldots\right)\nn\\
p^2{\cal F}_2&=&    p^2 \left(\frac{1}{120 a^2}+\ldots\right)
+p s^2 \left(-\frac{7}{720 a^2}-\frac{17}{384c^3}\ldots\right)+s^4
   \left(\frac{131}{1536
   c^3}+\ldots\right)
  \eea
  in perfect agreement with (\ref{fh1inst}) up to terms without $c$ or containing log functions\footnote{Since our prepotential is computed integrating twice
the coupling $\tau$ we do not control the polynomial terms of order two, while the log terms which come from perturbation theory on the gauge side are not present on the CFT side.}.
\subsection{${\cal H}_2$ theory}

Here we  repeat the exercise of the previous section for the ${\cal H}_2$ theory. The dynamics of the ${\cal H}_2$ theory is described by the elliptic curve
\bea
w^2 =-\left(M^2 \, z^4+ 2u \, z^3+m z^2 +c z+1\right)
\eea
 or equivalently by a curve in Weierstrass form with
 \bea
 g_2(u) &=& \ft43 (m^2+12 M^2-6 c u) \nn\\
 g_3 (u) &=& 4(4u^2+c^2 M^2) -\ft83 m(c u+4 M^2) +{8m^3\over 27} \label{g2g3h2}
 \eea
 For simplicity we will set $M^2=0$, $m=c^2/4$. For this choice the discriminant reduces to
 \be
 \Delta=1024 u^3 \left(c^3-108 u\right)
 \ee
Plugging (\ref{g2g3h2}) into (\ref{w1ee}), (\ref{f1h1}), (\ref{HAE}) one finds
\bea
&&-\ft12{\partial^2 {\cal F}_0 \over \partial a^2} = \log q \quad ,\qquad
p {\cal F}_1 ={p \over 4}  \log \left[ \frac{9 X(u)}{2E_2(q) } \right] +\ft{ s^2-2p }{24 }\log  \left[
1024 u^3 \left(c^3-108 u\right)\right] \nn\\
&& p^2 {\cal F}_2 = \frac{1}{32 u^2 \left(c^3-108 u\right)^2}  \left[\frac{405}{8} c^2 p^2X(u)^3+
X(u)^2\left(p^2 \left(\frac{81 c^4}{16}-1215 c u\right)+p s^2 \left(486 c u-\frac{27 c^4}{8}\right)\right)
\right. \nn\\
&&
+X(u)\left(p^2 \left(\frac{27 c^6}{32}-153 c^3 u+9720 u^2\right)
+p s^2 \left(-\frac{3 c^6}{4}+162 c^3 u-11664 u^2\right)\right.\nn\\
&&\left.+s^4 \left(\frac{3 c^6}{32}-27 c^3 u+1944 u^2\right)\right)
+p s^2 \left(-\frac{31 c^8}{480}+\frac{29 c^5 u}{2}-720 c^2 u^2\right)\nn\\
&& \left.
+p^2 \left(\frac{83 c^8}{2880}-\frac{27 c^5 u}{4}+348 c^2 u^2\right)
+s^4 \left(\frac{23 c^8}{960}-\frac{21 c^5 u}{4}+252 c^2 u^2\right)\right] \nn
\eea
     with
 \be
  X(u)=\frac{\left(c^6-144 c^3 u+3456 u^2\right)E_2 E_4 }{18 c\left(c^3-96 u\right) E_6 }
  \ee
    The $X$-independent part has been fixed  imposing the gap conditions
  \bea
  p^2 {\cal F}_2 & \underset{u\to  0}  \approx &  {3 \over a^2} \times \left( \frac{p^2}{240 }-\frac{7 s^2 p }{1440}  +   \frac{7 s^4}{5760  } \right) +O(a^0) \nn\\
   p^2 {\cal F}_2 & \underset{u\to c^3/108} \approx &  {1 \over a^2} \times \left( \frac{p^2}{240 }-\frac{7 s^2 p }{1440}  +   \frac{7 s^4}{5760  } \right) +O(a^0)
  \eea
  near the two degenerated points. The extra factor of three comes from the fact that $u= 0$ is a third order zero of the discriminant.

  \subsection*{Comparison against CFT}

     To compare against the CFT results one starts from (\ref{phi2h2hat})
     and  sets
    \bea
   \hat{c}_0  &=& \frac{m}{2}-\frac{c^2}{8}\quad , \quad
 \hat{c}_1 =\frac{c}{2}\quad, \quad \hat{c}_2 =1\quad, \quad \hat{v}=u+\ldots \label{dicth2}
\eea
 to bring the quadratic differential  to the standard ${\cal H}_2$ form
   \be
\phi_2   ={M^2 \over z^2} + {u\over z^3} +{m\over z^4}+ {c\over z^5}+ {1\over z^6}
\ee
Setting the dimension $[a]=1$ one finds the dimensions
 \be
[c]=\ft12 \quad, \quad [u]=\ft32 \quad\quad [M]=[m]=1 \quad , \quad [z] =-\ft12
 \ee
as expected for the ${\cal H}_2$ AD theory.

Using the map (\ref{map_CFT_H2}) and (\ref{dicth2}) in (\ref{prepotentiah2cft}) with further specification
$M=0$, $m=c^2/4$ one finds for the instanton prepotential
 \bea
&& {\cal F}^{CFT}_{{\rm inst},{\cal H}_2} = p \log Z^{CFT}_{{\rm inst},{\cal H}_2} =-\frac{12 a^3}{c^2} -\frac{105 a^4}{c^4}
-  \frac{1608 a^5}{c^6} +p \left(\frac{3 a^2}{c^4}+\frac{168 a^3}{c^6}\right)\nn \\
&&-s^2\left(\frac{3 a}{c^2}+\frac{111 a^2}{2 c^4}
+\frac{1500 a^3}{c^6}\right)+p s^2 \left(\frac{1}{4 c^4}+\frac{46 a}{c^6}\right)
-s^4 \left(\frac{25}{16c^4}+\frac{281 a}{2 c^6}\right)+\ldots
\label{fh2inst}
\eea
 We can compare these results against those obtained from the holomorphic anomaly equations in the limit
 where $c$ is large. Notice that in this limit
  the discriminant vanishes for $u\approx 0$ while $q\approx 0$. Expanding (\ref{w1ee}) around $q\approx 0$ and $u\approx 0$ one finds
  \bea
q(a) &=&\frac{2 \sqrt{2}\, a^{3/2}}{c^3}\left[1+\frac{36 a}{c^2}+\frac{1278 a^2}{c^4}+
\frac{46536 a^3}{c^6}+\frac{1735614 a^4}{c^8}
{+}\ldots \right] \nn\\
u(a)& =&   {c\, a\over 2} \left[1-\frac{6 a}{c^2}-\frac{48 a^2}{c^4}-\frac{840 a^3}{c^6}
-\frac{19296 a^4}{c^8}-\frac{512064 a^5}{c^{10}}{+}\ldots \label{quh2}
\right]
  \eea
    leading to
\bea
{\cal F}_0&=&-\frac{3 a^2}{2} \log \frac{2a}{c^2} +\frac{9 a^2}{4}
-\frac{12a^3}{c^2}-\frac{105 a^4}{ c^4}-\frac{1608 a^5}{c^6}+\ldots \nn\\
p{\cal F}_1&=&\left(\frac{s^2}{8}-\frac{p}{4}\right) \log (8a)
+\frac{1}{4} s^2 \log \left({c\over 2} \right)
+ p \left(\frac{3 a^2}{c^4}+\frac{168 a^3}{c^6}\right)
-s^2 \left(\frac{3 a}{c^2}+\frac{111 a^2}{2 c^4}+\frac{1500 a^3}{c^6}\right)+\ldots \nn\\
p^2{\cal F}_2&=& \frac{p^2}{80 a^2}+p s^2 \left( -\frac{7}{480 a^2}+\frac{1}{4 c^4}
+\frac{46a}{c^6}\right)+s^4\left(\frac{7}{1920 a^2}-\frac{25}{16 c^4}-\frac{281a}{2 c^6}\right)+\ldots
  \eea
in agreement with  (\ref{fh2inst}) keeping into account the observations at the end of Subsection 3.3.
\section{NS limit and WKB analysis}

A consistent application of the localization technique in the NS limit of the $\Omega$-background leads to the notion of deformed SW curve expressed as a difference equation
\cite{Poghossian:2010pn,Fucito:2011pn} (see also \cite{Mironov:2009dv} for an earlier approach).
By means of a Fourier transform the latter turns into a Schr\"odinger-like equation
\bea
\left(\hbar^2\frac{d^2}{dz^2}-\hat\phi_2(z)\right)\psi(z)=0
\label{schroedingereq}
\eea
where $\hbar=\epsilon_2$.
The potential $\hat\phi_2(z)$ defines the SW-differential as in (\ref{SW_diff_vs_phi_2}), while
the wave function can be interpreted as the partition function of a quiver gauge theory which
is the AGT dual  \cite{Alday:2009aq} of a 2d CFT conformal block with a degenerate field insertion at $z$.
To perform a WKB analysis we represent the "wave function" as
\bea
\psi(z)=e^{\frac{1}{\hbar} {\cal F}(z)}
\eea
Technically the second order ordinary differential equation is
obtained from the NS limit of the BPZ equation obtained by inserting a degenerated field of
level two into the  correlator of $N$ chiral primary fields. $\hat\phi_2(z)$, the (normalized)
expectation value of the holomorphic stress energy tensor $T(z)$, can be written as
\be
\hat\phi_2(z)=-p\sum_{i=1}^N \left[  { \Delta_{\alpha_i} \over (z-z_i)^2} +{d_i\over (z-z_i) } \right]   \label{phi2}
\ee
where $z_i$, $\Delta_{\alpha_i}$ are the insertion points and the conformal dimensions of primary fields.
The coefficients $d_i$'s are constrained by the requirement that infinity be a regular point, i.e. $\hat\phi_2(z) \underset{z\to \infty}{\sim} z^{-4}$,
or equivalently
\be
\sum_{i=1}^N d_i =\sum_{i=1}^N \left(  d_i z_i+\Delta_{\alpha_i} \right)  =\sum_{i=1}^N \left(  d_i z^2_i+2\Delta_{\alpha_i} z_i \right)=0 \label{cons3}
\ee
that can be solved for three of them, let us say $d_1,d_2,d_N$ in terms of the $d_i$'s remaining ones which can be conveniently parameterised as\footnote{The $d_i$ are derivatives with respect to $z_i$ and the rescaling is to let the $u_i$'s be defined according to \cite{Matone:1995rx,Flume:2004rp}. The other terms in (\ref{ansd}) come from the rescaling of the irregular state by $\prod_{j\neq i}(z_j-z_i)^{\alpha_i\alpha_j}$ to get a finite limit, following \cite{Gaiotto:2012sf}.}
\be
d_i = { v_{i-2} \over  \prod_{j=1,2,N}'(z_j-z_i) }   +\sum_{j\neq i} {2\alpha_i \alpha_j\over z_j-z_i}  \qquad  i=3,\ldots N-1 \label{ansd}
\ee
The $v_i$'s are the Coulomb branch parameters and the prime denotes the omission of factors involving $z_j=\infty$.
Plugging (\ref{ansd}) into (\ref{phi2}) and using (\ref{cons3}) one can check that $\phi_2(z)$ is finite in the limit where $(n+1)$-points, let us say
$\{ z_{N-n},\ldots z_N \}$,
are sent to zero, keeping finite the combinations
\be
c_s=\sum_{i=N-n}^N \alpha_{i} z_i^s \qquad  \qquad    s=0,\ldots n
\ee
These equations can be solved for the $\alpha_i$'s in terms of the $c_s$'s. Plugging this into (\ref{phi2}) and sending $\{ z_1,\ldots z_{N-n-1} \}$ to infinity,
and $\{ z_{N-n},\ldots z_N\}$ to zero one finds a finite formula for $\phi_2(z)$ as a function of $v_i$, $c_s$ and  $\{ \alpha_1,\ldots \alpha_{N-n-1} \}$.

\subsection{The quantum period}

Denoting
\bea
P(z)=\frac{{\cal F}'(z) }{\hbar}
\eea
one brings the wave equation to the first order differential form
\bea
P'(z)+P(z)^2-\frac{1}{\hbar^2}\hat\phi_2(z)=0 \label{difp}
\eea
In the semiclassical approximation, one looks for the solutions of $P(z)$ as a power series in the Plank's constant
\bea
P(z)=\sum_{n=-1}^\infty\hbar^nP_n(z)
\eea
Plugging this into (\ref{difp}), one reduces the differential equation to the
the recursion relation
\bea
P_{n+1}(z)=\frac{1}{2\sqrt{\hat\phi_2(z)}}\left(
\frac{d}{dz}\,P_n(z)+\sum _{m=0}^n P_m(z) P_{n-m}(z)\right)
\eea
that allow us to derive recursively higher order terms $P_n(z)$, starting from
\bea
P_{-1}(z)=\sqrt{\hat\phi_2(z)}
\eea
It is easy to see that $P_n(z)$ terms with $n$ an even integer are always total derivatives, hence their periods around closed cycles vanish. Thus only $P_n(z)$
with odd $n$ are relevant to the computation of the prepotential. For the first few terms one finds
\bea
P_1(z)&=&-\frac{5 \hat\phi_2 '^2+4 \hat{\phi_2}  \hat\phi_2 ''}{32 \hat\phi_2 ^{5/2}} \\
P_3(z) &=&-\frac{221 \hat\phi_2 '^2 \hat\phi_2 ''}{256 \hat\phi_2 ^{9/2}}-
\frac{1105 \hat\phi_2 '^4}{2048 \hat\phi_2 ^{11/2}}-\frac{7 \hat\phi_2 ^{(3)} \hat\phi_2 '}{32 \hat\phi_2 ^{7/2}}
-\frac{19 \hat\phi_2 ''^2}{128 \hat\phi_2 ^{7/2}}+\frac{\hat\phi_2 ^{(4)}}{32 \hat\phi_2 ^{5/2}}\nonumber\\
P_5(z) &=&+\frac{248475 \hat\phi_2 '^4 \hat\phi_2 ''}{16384 \hat\phi_2 ^{15/2}}
-\frac{34503 \hat\phi_2 '^2 \hat\phi_2 ''^2}{4096 \hat\phi_2 ^{13/2}}
+\frac{1391 \hat\phi_2 ^{(3)} \hat\phi_2 ' \hat\phi_2 ''}{512 \hat\phi_2 ^{11/2}}-\frac{414125 \hat\phi_2 '^6}{65536 \hat\phi_2 ^{17/2}}
-\frac{1055 \hat\phi_2 ^{(3)} \hat\phi_2 '^3}{256 \hat\phi_2 ^{13/2}}
\nonumber\\
&+& \frac{815 \hat\phi_2 ^{(4)} \hat\phi_2 '^2}{1024 \hat\phi_2 ^{11/2}}-\frac{27 \hat\phi_2 ^{(5)} \hat\phi_2 '}{256 \hat\phi_2 ^{9/2}}
-\frac{55 \hat\phi_2 ^{(4)} \hat\phi_2 ''}{256 \hat\phi_2 ^{9/2}}
+\frac{631 \hat\phi_2 ''^3}{1024 \hat\phi_2 ^{11/2}}
-\frac{69 \left(\hat\phi_2^{(3)}\right)^2}{512 \hat\phi_2^{9/2}}+\frac{\hat\phi_2^{(6)}}{128 \hat\phi_2^{7/2}}\nonumber\\
&&\qquad\qquad\qquad\qquad\qquad\qquad\vdots\nonumber
\eea
These expressions allow us
to calculate the $\epsilon$-corrections to the prepotential up to order $\epsilon_2^6$.
 We will focus on the rank one case. For this choice, $\hat\phi_2$ can be viewed as a function of a single quantum Coulomb branch parameter $\hat v$.
The quantum $a$-period can be expanded as
\bea
\label{a_expansion_gen}
a(\hat{v})=a_0(\hat{v})+\epsilon_2^2a_2(\hat{v})+\epsilon_2^4a_4(\hat{v})+\epsilon_2^6a_6(\hat{v})+O(\epsilon_2^8)
\eea
with
\bea
\label{a_n_general}
a_n(\hat{v})=\oint_{\gamma_A} P_{n-1}(z)\frac{dz}{2\pi i}
\eea
This expansion can be inverted to find the modulus $\hat{v}$ as a function of
the flat coordinate $a$. Indeed $\hat{v}(a)$ can be
represented as a power series
\bea
\hat{v}(a)=v_0(a)+\epsilon_2^2v_2(a)+\epsilon_2^4v_4(a)+\epsilon_2^6v_6(a)+\cdots
\label{uvsa}\eea
Inserting (\ref{uvsa}) in (\ref{a_expansion_gen}) and comparing both sides of the equality
we get
\bea
\label{u_246_general}
&&v_2(a)= -\frac{a_2}{a_0'}\,\,; \quad
v_4(a)= -\frac{a_2{}^2 a_0''}{2 a_0'{}^3}
+\frac{a_2 a_2'}{a_0'{}^2}
-\frac{a_4}{a_0'}\\
&&v_6(a) = {-}\frac{a_2{}^3 a_0''{}^2}{2 a_0'{}^5}
{+}\frac{3 a_2{}^2 a_2' a_0''}{2 a_0'{}^4}-\frac{a_2{}^2 a_2''}{2 a_0'{}^3}{-}
\frac{a_4 a_2 a_0''}{a_0'{}^3}
{+}\frac{a_0{}^{(3)} a_2{}^3}{6 a_0'{}^4}
{+}\frac{a_2 a_4'}{a_0'{}^2}
{-}\frac{a_2 a_2'{}^2}{a_0'{}^3}
{+}\frac{a_4 a_2'}{a_0'{}^2}
{-}\frac{a_6}{a_0'}\nonumber
\eea
where  all the $a_n$'s and their derivatives on the r.h.s. are evaluated at $v_0$ defined as  $a_0(v_0(a) )= a$.
\subsection{${\cal H}_1$ Argyres-Douglas theory in the NS limit}

The ${\cal H}_1$ theory is obtained from the $N=4$ point function colliding the four singularities at $z=0$.
First we set
\be
c_s = \sum_{i=1}^4 \alpha_i  z_i^s \qquad  {\rm with} \qquad  s=0,\ldots 3\label{civsai}
\ee
solve for the $\alpha_i$'s in favour of the $c_s$'s and substitute into (\ref{phi2}). Then we set
\be
d_3 =  -{v \over (z_1-z_3)(z_2-z_3)(z_4-z_3) }+ {2 \alpha_1 \alpha_3\over z_1-z_3}+ {2 \alpha_2 \alpha_3\over z_2-z_3} +
{2 \alpha_2 \alpha_4\over z_2-z_4}
\ee
and solve (\ref{cons3}) for $d_1,d_2,d_4$ in favour of $d_3$. Plugging these $d_i$'s and $\alpha_i$'s into (\ref{phi2}) and sending $z_i$ to zero one finds
\be
\phi_2(z) =\frac{c_3^2}{z^8}+\frac{2 c_2 c_3}{z^7}+\frac{c_2^2+2
	c_1 c_3}{z^6}+\frac{2 \left(c_3 \left(c_0-2
	Q\right)+c_1 c_2\right)}{z^5}
+\frac{ 2v}{z^4}
\ee
which coincides with (\ref{phi2h1}), (\ref{vh1}).
Passing to the hatted variables using (\ref{map_CFT_H1}) we arrive to an
explicit expression (\ref{phi2hath1}) for $\hat\phi_2(z)$.
For small $\hat{c}_3$ the $A$-cycle again shrinks to a small contour around $z=0$ and the
integrals (\ref{a_n_general}) can be computed by means of residues. In this case we have expanded
the relevant quantities up to order $\hat{c}_3^8$ and computed their $\epsilon_2$ corrections.
Then using (\ref{u_246_general}) we have found $\hat{v}(a)$ up to corrections $\epsilon_2^6$.
The results of our computations are displayed in appendix \ref{WKB_results_h1}.

$\hat{v}(a)$ can be alternatively found from the irregular state computation
by plugging  (\ref{prepotentiah1cft}) into (\ref{vh1}), using (\ref{dicth1}),  (\ref{map_CFT_H1})  to pass  to
the hatted variables and finally specifying $\epsilon_1=0$.
We have checked that both methods are in perfect agreement.

\subsection{${\cal H}_2$ Argyres-Douglas theory in the NS limit}

The ${\cal H}_2$ theory is obtained from the $N=4$ point function colliding $n=3$ singularities at $z=0$ and one at infinity.
First we set
\be
c_s = \sum_{i=2}^4 \alpha_i  z_i^s \qquad  {\rm with} \qquad  s=0,\ldots 2\nn\\
\ee
and solve for $\alpha_2,\alpha_3,\alpha_4$ in favor of the $c_s$'s. Then we set
\be
d_3 = - {v_1 \over (z_2-z_3)(z_4-z_3) }+ {2 \alpha_2 \alpha_3\over z_2-z_3} +
{2 \alpha_2 \alpha_4\over z_2-z_4}  \nn\\
\ee
and solve (\ref{cons3}) for $d_1,d_2,d_4$ in favor of $d_3$. Plugging the $d_i$'s and the $\alpha_i$'s into (\ref{phi2}) and sending $z_1$ to infinity and
$z_2,z_3,z_4$ to zero one finds
\be
\phi_2(z)=\frac{c_2^2}{z^6}+\frac{2 c_1 c_2}{z^5}+\frac{c_2 \left(2c_0-3Q\right)+c_1^2}{z^4}
+\frac{2v}{z^3}-{\Delta_{\alpha_1} \over z^2}
\ee
which is the same as (\ref{phi2h2}), (\ref{vh2}) if one identifies
\be
\label{u_1_vs_LogZh2}
\alpha_1\equiv \beta_0\,,\qquad v=\frac{v_1}{2}+ c_1 \left(c_0-Q\right)
\ee
Thus using an alternative road we have re-derived (\ref{phi2h2hat}) which gives the $\phi_2(z)$ needed for the WKB analysis.
After going to hatted variables, it is easy to see that at small $\hat c_2$ the $A$-cycle can be chosen as a tiny contour
surrounding $z=0$, and the integrals (\ref{a_n_general}) simply pick up the residue at $z=0$.
We have computed this residue for $a_{0,2,4,6}$ up to order $\hat c_2^6$, then found $\hat{v}(a)$,
using (\ref{u_246_general}). The final expressions are given in \ref{WKB_results_h2}.
As in the ${\cal H}_1$ case, these expressions are in complete agreement with the CFT result obtained
from (\ref{u_1_vs_LogZh2}), (\ref{prepotentiah2cft}), (\ref{Zh2tree}), (\ref{vh2})
and the map (\ref{map_CFT_H2})
in the NS limit $\epsilon_1=0$.

\section{Conclusions}

In this paper we have carried out a comparison between the non perturbative results obtained for a ${\cal N}=2$ SCFT of the AD type with the recursion relations coming from a conformal anomaly \cite{Bershadsky:1993ta,Huang:2006si,Huang:2009md,Huang:2011qx,Huang:2013eja,Krefl:2010fm,Billo:2013fi,Billo:2013jba,Billo:2015pjb,Billo:2015jyt} and those obtained from the AGT duality for irregular states \cite{Bonelli:2011aa,Gaiotto:2012sf}, finding an agreement also keeping into account gravitational corrections. This comparison is possible in  a region away from the conformal point and holds also in the NS limit which has been recently related to scalar waves propagating in black holes and fuzzball geometries \cite{Bonelli:2021uvf,Bonelli:2022ten,Consoli:2022eey}.
 In this framework the parameters of the gravity solution: mass, charge, angular momenta, wave frequency codify masses and couplings of the gauge theory. In particular the radial propagation of scalar waves in asymptotically AdS/dS Kerr-Newman black holes is described by
a second order ordinary differential equation of the type of (\ref{schroedingereq}) with four Fuchsian singularities located at the four horizons, i.e. the zeros of
\be
\Delta(r)=(r^2+a^2)\left(1+ \epsilon {r^2\over L^2} \right) -2 M\,r +Q^2
\label{KNhorizon}
\ee
with $\epsilon$ equal $1$, $-1$ and $0$ for asymptotically AdS, dS and flat spaces respectively.
 It is  natural to ask whether a black hole exists such that its dynamics is described by an AD theory.
The AD points in the gauge theory correspond to points where three or more singularities collide. This can be achieved by taking De Sitter space ($\epsilon=-1$) and choosing mass, charge and angular momentum such that three singularities collide, i.e.  $\Delta(r)=(r-A)^3(r+3 A)$ for some $A$.
The solution depends on a single parameter $A$, to which mass, charge and angular momentum are related via
\bea
a=\sqrt{L^2-6A^2}\qquad, \qquad
M=\frac{4 A^3}{L^2}\qquad , \qquad
Q=\frac{\sqrt{3 A^4 + 6 A^2 L^2 - L^4}}{L}
\eea
 The solution exists if $ \ft{2}{\sqrt{3}}-1 < A^2/L^2 < 1/6$.
It is then easy to get convinced that it is possible to obtain $a>0, Q>0, M>0$ for many values of $L$. The solution of the scalar perturbations in this geometry are then directly related to the partition function of the SYM theory in the NS limit for the AD theory ${\cal H}_2$ which we computed in Section 4.
\acknowledgments
R.Poghossian thanks the INFN Roma Tor Vergata and the Armenian SCS
grant 21AG-1C062 for financial support.
The work of F.Fucito and J.F.Morales is partially supported by the MIUR PRIN Grant 2020KR4KN2 "String Theory as a bridge between Gauge Theories and Quantum Gravity".


  \begin{appendix}

\section{Prepotential from irregular states}

In this section we collect the results for the irregular conformal blocks
\bea
 Z_{{\cal H}_2} &=&  \langle \beta_0  |I_2({\bf c}, {\bm \beta})\rangle \nn\\
 Z_{{\cal H}_1} &=& \langle  0  |I_3({\bf c}, {\bm \beta})\rangle
\eea
as an expansion around the monopole point.

\subsection{ ${\cal H}_2$ theory}
\begin{small}
 \bea
&& \ln Z_{ {\cal H}_2 \rm inst}=\nonumber\\&&\frac{c_2}{2 c_1^2}[(3 Q+c_0 -3 \beta_2) (Q-\beta_2) (c_0
 -\beta_2)+(2 c_0 -3 \beta_2) (\beta_2-\beta_0) (-Q+\beta_0+\beta_2)]+\frac{c_2^2}{16 c_1^4}\left\{ \right.\nonumber\\
&&42 Q^3
(\beta_2-c_0 )+\left.\beta_0^2\left(16 c_0 ^2-72 \beta_2 c_0 +66 \beta_2^2-1\right)-\beta_0^4 -Q^2 \left[30 c_0 ^2-178 \beta_2 c_0 +\beta_0^2+147 \beta_2^2\right]+\right.\nonumber\\ 	
&+&\left.24 \beta_0 (c_0-\beta_2)+\beta_2 (4 c_0 ^3-48 \beta_2 c_0 ^2+2 c_0 (70 \beta_2^2-1)  -105 \beta_2^3+3 \beta_2)+ Q[\beta_0-3 \beta_2+\right.\nonumber\\
&+&\left.2 (-2 c_0 ^3+39 \beta_2 c_0 ^2-138 \beta_2^2 c_0+c_0 +\beta_0^3+105 \beta_2^3+12 \beta_0^2 (c_0 -\beta_2)+\beta_0 (-8 c_0 ^2+36 \beta_2
c_0 -33 \beta_2^2 ))]\right\}+\nonumber\\
&+&\frac{c_2^3}{48c_1^6}\{ 414 (c_0 -\beta_2) Q^4+ 3 Q^3[152 c_0 ^2-766 \beta_2 c_0 -\beta_0^2+615 \beta_2^2+84\beta_0 (c_0 -\beta_2)]
+\nonumber\\
&+&Q^2[132 c_0 ^3+4 c_0 ^2(63 \beta_0-418 \beta_2) +c_0(-218 \beta_0^2-996 \beta_2 \beta_0+4794 \beta_2^2-66)  +3 (2 \beta_0^3+246 \beta_2^2 \beta_0+\nonumber\\
&+&\beta_0-1080 \beta_2^3+\beta_2(68 \beta_0^2+21))]+Q[10 c_0 ^4+c_0 ^3(64 \beta_0-324
\beta_2) +4 c_0 ^2(-63 \beta_0^2-135 \beta_2 \beta_0+564 \beta_2^2-5) +\nonumber\\
&+&2 c_0(-34
\beta_0^3+498 \beta_2 \beta_0^2+654 \beta_2^2 \beta_0-23 \beta_0-2460 \beta_2^3+96 \beta_2)+3015 \beta_2^4-900 \beta_0 \beta_2^3-9 \beta_2^2(82 \beta_0^2+21)+\nonumber\\
&+&\beta_2(96 \beta_0^3+66 \beta_0)-3(\beta_0^4+\beta_0^2)]+2 \beta_0^4 (17c_0 -24 \beta_2)-2 \beta_2[ 603 \beta_2^4-1005 c_0  \beta_2^3+(520 c_0 ^2-63) \beta_2^2+\nonumber\\
&-&96
c_0 ^3 \beta_2+63 c_0  \beta_2+5 c_0 ^2 (c_0 ^2-2)]+2 \beta_0^2 (-32 c_0 ^3+270 \beta_2c_0 ^2+(23-654 \beta_2^2) c_0 +450 \beta_2^3-33 \beta_2)\}
\label{prepotentiah2cft}\eea
\end{small}
 \subsection{ ${\cal H}_1$ theory}
\begin{small}
 \bea
&& \ln Z_{ {\cal H}_1 \rm inst}=\nonumber\\&-&\frac{c_1 c_3}{3 c_2^3}\left[2 c_1^2 (-2 \beta_2+c_0+Q)+3 c_2 (6 \beta_2^2-6 \beta_2 (c_0+Q)+c_0
 (c_0+5 Q))\right]+\frac{c_1c_3^3}{6 c_2^7}\left\{\right.\nonumber\\
 &-& \left. 3 c_2^2 (-2 \beta_2+c_0+Q) (34 \beta_2^2-1+4 c_0^2-34 \beta_2 (c_0+Q)+35c_0Q)-12 c_1^4 (-2 \beta_2+c_0+Q)\right.\nonumber\\
 &-&\left.32 c_1^2 c_2 \left[6 \beta_2^2-6 \beta_2(c_0+Q)+c_0 (c_0+5 Q)\right]\right\}-\frac{c_3^2 }{12 c_2^5}\left\{c_2^2 (-2 \beta_2+c_0+Q) (34 \beta_2^2+4 c_0^2\right.\nonumber\\
 &-&\left.34 \beta_2 (c_0+Q)+35 c_0 Q-1)+12 c_1^4 (-2 \beta_2+c_0+Q)+24 c_1^2 c_2 \left[6 \beta_2^2-6 \beta_2 (c_0+Q)+\right.\right.\nonumber\\
 &+&\left.\left.c_0(c_0+5 Q)\right]\right\}+\frac{c_3^4}{24 c_2^9}\left\{-60 c_1^2 c_2^2 (-2 \beta_2+c_0+Q) (34 \beta_2^2+4 c_0^2-34 \beta_2 (c_0+Q)+35c_0 Q-1)+\right.\nonumber\\
 &-&\left.112 c_1^6 (-2 \beta_2+c_0+Q)-384 c_1^4 c_2 [6 \beta_2^2-6 \beta_2
 (c_0+Q)+c_0 (c_0+5 Q)]+c_2^3 [-750 \beta_2^4-16 c_0^4+\right.\nonumber\\
 &-&\left.241 c_0^3 Q-6 \beta_2^2 (166 c_0^2+461 c_0 Q+178 Q^2-15)+6 \beta_2
 (c_0+Q) (41 c_0^2+211 c_0 Q+53 Q^2-15)+\right.\nonumber\\
 &-&\left.508 c_0^2
 Q^2+14 c_0^2-315 c_0 Q^3+1500 \beta_2^3 (c_0+Q)+79 c_0
 Q-3 Q^2]\right\}+\frac{c_1c_3^5}{30 c_2^{11}}  [-360 c_1^6 (-2 \beta_2+c_0+Q)+\nonumber\\
 &-&1536 c_1^4 c_2 [6 \beta_2^2-6 \beta_2
 (c_0+Q)+c_0 (c_0+5 Q)]-350 c_1^2 c_2^2 (-2 \beta_2+c_0+Q)(34 \beta_2^2+4 c_0^2-34 \beta_2
 (c_0+Q)+\nonumber\\
&+&35 c_0 Q-1)+15 c_2^3 (-16c_0^4-241 c_0^3 Q-508 c_0^2 Q^2+14 c_0^2-315 c_0 Q^3+79 c_0 Q-3 Q^2)+\nonumber\\
&+&1500 \beta_2^3 (c_0+Q)-750\beta_2^4
+6 \beta_2 (c_0+Q) (41 c_0^2+211 c_0
 Q+53 Q^2-15)-6 \beta_2^2 (166 c_0^2+461 c_0
 Q+\nonumber\\
 &+&178 Q^2-15)]
\label{prepotentiah1cft}\eea
\end{small}
\section{Results in the NS limit}
\subsection{${\cal H}_1$ theory}
\label{WKB_results_h1}
The quadratic differential $\phi_2dz^2$ in this case is given by (\ref{phi2hath1}).
 The integrals (\ref{a_n_general}) in the small $c_3$
limit can be computed by taking the residue at $z=0$. For un-deformed $a_0$ we got
\begin{small}
\bea
\label{a0H1}
a_0(\hat v)&=&\frac{2 \hat v- \hat c_1^2}{2  \hat c_2}- \hat c_3 (\frac{ \hat c_1( \hat c_1^2-2 \hat v)}{ \hat c_2^3}
+\frac{ \hat c_0  \hat c_1}{ \hat c_2^2})\nonumber\\
&-& \hat c_3^2 \left(\frac{3  \hat c_0 \left(3  \hat c_1^2-2 \hat v\right)}{2  \hat c_2^4}
+\frac{3 \left( \hat c_1^2-2 \hat v\right) \left(5  \hat c_1^2-2 \hat v\right)}{4  \hat c_2^5}
+\frac{ \hat c_0^2}{2  \hat c_2^3}\right)\nonumber\\
&-& \hat c_3^3 \left(\frac{5  \hat c_1  \hat c_0 \left(5  \hat c_1^2-6 \hat v\right)}{ \hat c_2^6}
+\frac{5  \hat c_1 \left( \hat c_1^2-2 \hat v\right) \left(7  \hat c_1^2-6 \hat v\right)}{2  \hat c_2^7}
+\frac{6  \hat c_1  \hat c_0^2}{ \hat c_2^5}\right)\nonumber\\
&-& \hat c_3^4 \left(\frac{35  \hat c_0 \left(-60  \hat c_1^2 \hat v+35  \hat c_1^4+12 \hat v^2\right)}{8  \hat c_2^8}
+\frac{35 \left( \hat c_1^2-2 \hat v\right) \left(-28  \hat c_1^2 \hat v+21  \hat c_1^4+4 \hat v^2\right)}{8  \hat c_2^9}\right.\nn\\
&&\left.+\frac{45  \hat c_0^2 \left(5  \hat c_1^2-2 \hat v\right)}{4  \hat c_2^7}+\frac{5  \hat c_0^3}{2  \hat c_2^6}\right)\nonumber\\
&-& \hat c_3^5 \left(\frac{63  \hat c_1  \hat c_0 \left(-140  \hat c_1^2 \hat v+63  \hat c_1^4+60 \hat v^2\right)}{4  \hat c_2^{10}}
+\frac{63  \hat c_1 \left( \hat c_1^2-2 \hat v\right) \left(-60  \hat c_1^2 \hat v+33  \hat c_1^4+20 \hat v^2\right)}{4  \hat c_2^{11}}\right.\nn\\
&&\left.+\frac{70  \hat c_1  \hat c_0^2 \left(7  \hat c_1^2-6 \hat v\right)}{ \hat c_2^9}+\frac{105  \hat c_1  \hat c_0^3}{2  \hat c_2^8}\right)\nonumber\\
&-& \hat c_3^6 \left(\frac{1575  \hat c_0^2 \left(-28  \hat c_1^2 \hat v+21  \hat c_1^4+4 \hat v^2\right)}{8  \hat c_2^{11}}
+\frac{231  \hat c_0 \left(420  \hat c_1^2 \hat v^2-630  \hat c_1^4 \hat v+231  \hat c_1^6-40 \hat v^3\right)}{8  \hat c_2^{12}}\right.\nn\\
&&\left.+\frac{231 \left( \hat c_1^2-2 \hat v\right) \left(540  \hat c_1^2 \hat v^2-990  \hat c_1^4 \hat v+429  \hat c_1^6-40 \hat v^3\right)}{32  \hat c_2^{13}}
+\frac{105  \hat c_0^3 \left(7  \hat c_1^2-2 \hat v\right)}{ \hat c_2^{10}}+\frac{35  \hat c_0^4}{2  \hat c_2^9}\right)\nonumber\\
&-& \hat c_3^7 \left(\frac{2079  \hat c_1  \hat c_0^2 \left(-60  \hat c_1^2 \hat v+33  \hat c_1^4+20 \hat v^2\right)}{2  \hat c_2^{13}}
+\frac{429  \hat c_1  \hat c_0 \left(1260  \hat c_1^2 \hat v^2-1386  \hat c_1^4 \hat v+429  \hat c_1^6-280 \hat v^3\right)}{4  \hat c_2^{14}}\right.\nn\\
&&+\frac{429  \hat c_1 \left( \hat c_1^2-2 \hat v\right) \left(1540  \hat c_1^2 \hat v^2-2002  \hat c_1^4 \hat v
	+715  \hat c_1^6-280 \hat v^3\right)}{16  \hat c_2^{15}}\nn\\
&&\left.+\frac{5775  \hat c_1  \hat c_0^3 \left(3  \hat c_1^2-2 \hat v\right)}{2  \hat c_2^{12}}+\frac{525  \hat c_1  \hat c_0^4}{ \hat c_2^{11}}\right)\nonumber\\
&-& \hat c_3^8 \left(\frac{45045  \hat c_0^3 \left(-36  \hat c_1^2 \hat v+33  \hat c_1^4+4 \hat v^2\right)}{16  \hat c_2^{14}}
+\frac{21021  \hat c_0^2 \left(540  \hat c_1^2 \hat v^2-990  \hat c_1^4 \hat v+429  \hat c_1^6-40 \hat v^3\right)}{32  \hat c_2^{15}}\right.\nn\\
&&+\frac{6435  \hat c_0 \left(27720  \hat c_1^4 \hat v^2-10080  \hat c_1^2 \hat v^3-24024  \hat c_1^6 \hat v+6435  \hat c_1^8+560 \hat v^4\right)}{128  \hat c_2^{16}}\nn\\
&&+\frac{6435 \left( \hat c_1^2-2 \hat v\right) \left(8008  \hat c_1^4 \hat v^2-2464  \hat c_1^2 \hat v^3-8008  \hat c_1^6 \hat v+2431  \hat c_1^8
	+112 \hat v^4\right)}{128  \hat c_2^{17}}\nn\\
&&\left.+\frac{17325  \hat c_0^4 \left(9  \hat c_1^2-2 \hat v\right)}{16  \hat c_2^{13}}
+\frac{1155  \hat c_0^5}{8  \hat c_2^{12}}\right)+O( \hat c_3^9)
\eea
\end{small}
and for $A$-cycle corrections $a_{2,4,6}$ we obtained
\begin{small}
\bea
a_2(\hat v)&=&-\frac{ \hat c_3^2}{4  \hat c_2^3}-\frac{3  \hat c_1  \hat c_3^3}{2  \hat c_2^5}
- \hat c_3^4 \left(\frac{5 \left(17  \hat c_1^2-10 \hat v\right)}{8  \hat c_2^7}+\frac{25  \hat c_0}{8  \hat c_2^6}\right)
- \hat c_3^5 \left(\frac{35  \hat c_1 \left(9  \hat c_1^2-10 \hat v\right)}{4  \hat c_2^9}
+\frac{175  \hat c_0  \hat c_1}{4  \hat c_2^8}\right)\nonumber\\
&-& \hat c_3^6 \left(\frac{105 \left(-148  \hat c_1^2 \hat v+91  \hat c_1^4+28 \hat v^2\right)}{16  \hat c_2^{11}}
+\frac{105  \hat c_0 \left(37  \hat c_1^2-14 \hat v\right)}{8  \hat c_2^{10}}+\frac{315  \hat c_0^2}{8  \hat c_2^9}\right)\nonumber\\
&-& \hat c_3^7 \left(\frac{231  \hat c_1 \left(-340  \hat c_1^2 \hat v+159  \hat c_1^4+140 \hat v^2\right)}{8  \hat c_2^{13}}
+\frac{1155  \hat c_1  \hat c_0 \left(17  \hat c_1^2-14 \hat v\right)}{4  \hat c_2^{12}}+\frac{1785  \hat c_1  \hat c_0^2}{2  \hat c_2^{11}}\right)\nonumber\\
&-& \hat c_3^8 \left(\frac{15015  \hat c_0 \left(-260  \hat c_1^2 \hat v+201  \hat c_1^4+36 \hat v^2\right)}{64  \hat c_2^{14}}
+\frac{3003 \left(1300  \hat c_1^2 \hat v^2-2010  \hat c_1^4 \hat v+759  \hat c_1^6-120 \hat v^3\right)}{64  \hat c_2^{15}}\right.\nn\\
&&\left.+\frac{1155  \hat c_0^2 \left(381  \hat c_1^2-106 \hat v\right)}{32  \hat c_2^{13}}
+\frac{8085  \hat c_0^3}{16  \hat c_2^{12}}\right)+O( \hat c_3^9)\nonumber\\
a_4(\hat v)&=&-\frac{175  \hat c_3^6}{32  \hat c_2^9}-\frac{1575  \hat c_1  \hat c_3^7}{16  \hat c_2^{11}}-
 \hat c_3^8 \left(\frac{3465 \left(49  \hat c_1^2-18 \hat v\right)}{128  \hat c_2^{13}}
+\frac{31185  \hat c_0}{128  \hat c_2^{12}}\right)+O( \hat c_3^9)\nonumber\\
a_6(\hat v)&=&O( \hat c_3^9)
\eea
\end{small}
The inverse of $a_0(\hat v)$ (\ref{a0H1}) reads
\begin{small}
\bea
&& v_0(a)=a  \hat c_2+\frac{ \hat c_1^2}{2}+\frac{ \hat c_1  \hat c_3 \left( \hat c_0-2 a\right)}{ \hat c_2}
+ \hat c_3^2 \left(\frac{6 a^2-6 a  \hat c_0+ \hat c_0^2}{2  \hat c_2^2}+\frac{ \hat c_1^2 \left( \hat c_0-2 a\right)}{ \hat c_2^3}\right)\nonumber\\
&&+ \hat c_3^3 \left(\frac{2  \hat c_1 \left(6 a^2-6 a  \hat c_0+ \hat c_0^2\right)}{ \hat c_2^4}
+\frac{2  \hat c_1^3 \left( \hat c_0-2 a\right)}{ \hat c_2^5}\right)\nonumber\\
&&+ \hat c_3^4 \left(\frac{8  \hat c_1^2 \left(6 a^2-6 a  \hat c_0+ \hat c_0^2\right)}{ \hat c_2^6}
+\frac{\left( \hat c_0-2 a\right) \left(17 a^2-17 a  \hat c_0+2  \hat c_0^2\right)}{2  \hat c_2^5}
+\frac{5  \hat c_1^4 \left( \hat c_0-2 a\right)}{ \hat c_2^7}\right)\nonumber\\
&&+ \hat c_3^5 \left(\frac{32  \hat c_1^3 \left(6 a^2-6 a  \hat c_0+ \hat c_0^2\right)}{ \hat c_2^8}
+\frac{5  \hat c_1 \left( \hat c_0-2 a\right) \left(17 a^2-17 a  \hat c_0+2  \hat c_0^2\right)}{ \hat c_2^7}
+\frac{14  \hat c_1^5 \left( \hat c_0-2 a\right)}{ \hat c_2^9}\right)\nonumber\\
&&+ \hat c_3^6 \left(\frac{128  \hat c_1^4 \left(6 a^2-6 a  \hat c_0+ \hat c_0^2\right)}{ \hat c_2^{10}}
+\frac{35  \hat c_1^2 \left( \hat c_0-2 a\right) \left(17 a^2-17 a  \hat c_0+2  \hat c_0^2\right)}{ \hat c_2^9}\right.\nn\\
&&\left.+\frac{-750 a^3  \hat c_0+498 a^2  \hat c_0^2+375 a^4-123 a  \hat c_0^3+8  \hat c_0^4}{2  \hat c_2^8}
+\frac{42  \hat c_1^6 \left( \hat c_0-2 a\right)}{ \hat c_2^{11}}\right)\nonumber\\
&&+ \hat c_3^7 \left(\frac{512  \hat c_1^5 \left(6 a^2-6 a  \hat c_0+ \hat c_0^2\right)}{ \hat c_2^{12}}
+\frac{210  \hat c_1^3 \left( \hat c_0-2 a\right) \left(17 a^2-17 a  \hat c_0+2  \hat c_0^2\right)}{ \hat c_2^{11}}\right.\nn\\
&&\left.+\frac{8  \hat c_1 \left(-750 a^3  \hat c_0+498 a^2  \hat c_0^2+375 a^4-123 a  \hat c_0^3+8  \hat c_0^4\right)}{ \hat c_2^{10}}
+\frac{132  \hat c_1^7 \left( \hat c_0-2 a\right)}{ \hat c_2^{13}}\right)\nonumber\\
&&+ \hat c_3^8 \left(\frac{2048  \hat c_1^6 \left(6 a^2-6 a  \hat c_0+ \hat c_0^2\right)}{ \hat c_2^{14}}
+\frac{1155  \hat c_1^4 \left( \hat c_0-2 a\right) \left(17 a^2-17 a  \hat c_0+2  \hat c_0^2\right)}{ \hat c_2^{13}}\right.\nn\\
&&+\frac{80  \hat c_1^2 \left(-750 a^3  \hat c_0+498 a^2  \hat c_0^2+375 a^4-123 a  \hat c_0^3+8  \hat c_0^4\right)}{ \hat c_2^{12}}\nn\\
&&\left.+\frac{3 \left( \hat c_0-2 a\right) \left(-7126 a^3  \hat c_0+4602 a^2  \hat c_0^2+3563 a^4-1039 a  \hat c_0^3+56  \hat c_0^4\right)}
{8  \hat c_2^{11}}+\frac{429  \hat c_1^8 \left( \hat c_0-2 a\right)}{ \hat c_2^{15}}\right)\nn\\
\eea
\end{small}
Using formulae  (\ref{u_246_general}), for the $\epsilon_2$ corrections $ v_{2,4,6}$ we obtain
\begin{small}
\bea
&& v_2(a)=\frac{ \hat c_3^2}{4  \hat c_2^2}+\frac{ \hat c_1  \hat c_3^3}{ \hat c_2^4}+ \hat c_3^4
\left(\frac{19 \left( \hat c_0-2 a\right)}{8  \hat c_2^5}+\frac{4  \hat c_1^2}{ \hat c_2^6}\right)
+ \hat c_3^5 \left(\frac{95  \hat c_1 \left( \hat c_0-2 a\right)}{4  \hat c_2^7}+\frac{16  \hat c_1^3}{ \hat c_2^8}\right)\nonumber\\
&&+ \hat c_3^6 \left(\frac{459 a^2-459 a  \hat c_0+97  \hat c_0^2}{4  \hat c_2^8}
+\frac{665  \hat c_1^2 \left( \hat c_0-2 a\right)}{4  \hat c_2^9}
+\frac{64  \hat c_1^4}{ \hat c_2^{10}}\right)\nonumber\\
&&+ \hat c_3^7 \left(\frac{4  \hat c_1 \left(459 a^2-459 a  \hat c_0+97  \hat c_0^2\right)}{ \hat c_2^{10}}
+\frac{1995  \hat c_1^3 \left( \hat c_0-2 a\right)}{2  \hat c_2^{11}}
+\frac{256  \hat c_1^5}{ \hat c_2^{12}}\right)\nonumber\\
&&+ \hat c_3^8 \left(\frac{40  \hat c_1^2 \left(459 a^2-459 a  \hat c_0+97  \hat c_0^2\right)}{ \hat c_2^{12}}
+\frac{\left( \hat c_0-2 a\right) \left(23405 a^2-23405 a  \hat c_0+4004  \hat c_0^2\right)}{16  \hat c_2^{11}}\right.\nn\\
&&\left.+\frac{21945  \hat c_1^4 \left( \hat c_0-2 a\right)}{4  \hat c_2^{13}}
+\frac{1024  \hat c_1^6}{ \hat c_2^{14}}\right)+ O( \hat c_3^9)\nonumber\\
&& v_4(a)=\frac{131  \hat c_3^6}{32  \hat c_2^8}+\frac{131  \hat c_1  \hat c_3^7}{2  \hat c_2^{10}}
+ \hat c_3^8 \left(\frac{22709 \left( \hat c_0-2 a\right)}{128  \hat c_2^{11}}
+\frac{655  \hat c_1^2}{ \hat c_2^{12}}\right)+O( \hat c_3^9)\nonumber\\
&& v_6(a)=O( \hat c_3^9)
\eea
\end{small}
\subsection{${\cal H}_2$ theory}
\label{WKB_results_h2}
The quadratic differential is the $\hat\phi_2dz^2$ in (\ref{phi2h2hat}).
The integrals (\ref{a_n_general}) in the small $ \hat c_2$ limit are computed by taking the residue at $z=0$.
We have kept all terms up to the order $ \hat c_2^6$. For un-deformed $a_0$ we find
\begin{small}
\bea
\label{a0H2}
&&a_0( \hat v)=\frac{ \hat v}{ \hat c_1}+ \hat c_2 \left(-\frac{M^2}{2  \hat c_1^2}
+\frac{3  \hat v^2}{2  \hat c_1^4}-\frac{ \hat c_0  \hat v}{ \hat c_1^3}\right)+
 \hat c_2^2 \left(\frac{3  \hat v \left( \hat c_0^2-2 M^2\right)}{2  \hat c_1^5}+\frac{3  \hat c_0 M^2}{2  \hat c_1^4}+\frac{15  \hat v^3}{2  \hat c_1^7}-\frac{15  \hat c_0  \hat v^2}{2  \hat c_1^6}\right)\nonumber\\
&&+ \hat c_2^3 \left(\frac{105  \hat v^2 \left( \hat c_0^2-M^2\right)}{4  \hat c_1^8}
-\frac{5  \hat c_0  \hat v \left( \hat c_0^2-9  \hat c_2M^2\right)}{2  \hat c_1^7}-\frac{5 M^2 \left(3  \hat c_0^2-M^2\right)}{4  \hat c_1^6}
+\frac{105  \hat v^4}{2  \hat c_1^{10}}-\frac{70  \hat c_0  \hat v^3}{ \hat c_1^9}\right)\nonumber\\
&&+ \hat c_2^4 \left(\frac{35  \hat v \left(-24  \hat c_0^2 M^2+ \hat c_0^4+6 M^4\right)}{8  \hat c_1^9}
+\frac{525  \hat v^3 \left(3  \hat c_0^2-2 M^2\right)}{4  \hat c_1^{11}}\right.\nn\\
&&\left.-\frac{315  \hat c_0  \hat v^2 \left( \hat c_0^2-4 M^2\right)}{4  \hat c_1^{10}}
+\frac{35  \hat c_0 M^2 \left(2  \hat c_0^2-3 M^2\right)}{8  \hat c_1^8}
+\frac{3465  \hat v^5}{8  \hat c_1^{13}}-\frac{5775  \hat c_0  \hat v^4}{8  \hat c_1^{12}}\right)\nn\\
&&+ \hat c_2^5 \left(\frac{3465  \hat v^2 \left(-10  \hat c_0^2 M^2+ \hat c_0^4+2 M^4\right)}{16  \hat c_1^{12}}
-\frac{63  \hat c_0  \hat v \left(-50  \hat c_0^2 M^2+ \hat c_0^4+50 M^4\right)}{8  \hat c_1^{11}}\right.\nn\\
&&-\frac{63 M^2 \left(-20  \hat c_0^2 M^2+5  \hat c_0^4+2 M^4\right)}{16  \hat c_1^{10}}
+\frac{45045  \hat v^4 \left(2  \hat c_0^2-M^2\right)}{16  \hat c_1^{14}}\nn\\
&&\left.-\frac{3465  \hat c_0  \hat v^3 \left(2  \hat c_0^2-5 M^2\right)}{4  \hat c_1^{13}}
+\frac{63063  \hat v^6}{16  \hat c_1^{16}}-\frac{63063  \hat c_0  \hat v^5}{8  \hat c_1^{15}}\right)\nn\\
&&+ \hat c_2^6 \left(\frac{105105  \hat v^3 \left(-6  \hat c_0^2 M^2+ \hat c_0^4+M^4\right)}{16  \hat c_1^{15}}
-\frac{9009  \hat c_0  \hat v^2 \left(-20  \hat c_0^2 M^2+ \hat c_0^4+15 M^4\right)}{16  \hat c_1^{14}}\right.\nn\\
&&+\frac{231  \hat v \left(225  \hat c_0^2 M^4-90  \hat c_0^4 M^2+ \hat c_0^6-20 M^6\right)}{16  \hat c_1^{13}}
+\frac{231  \hat c_0 M^2 \left(-25  \hat c_0^2 M^2+3  \hat c_0^4+10 M^4\right)}{16  \hat c_1^{12}}\nn\\
&&\left.+\frac{63063  \hat v^5 \left(5  \hat c_0^2-2 M^2\right)}{4  \hat c_1^{17}}
-\frac{105105  \hat c_0  \hat v^4 \left(5  \hat c_0^2-9 M^2\right)}{16  \hat c_1^{16}}
+\frac{153153  \hat v^7}{4  \hat c_1^{19}}-\frac{357357  \hat c_0  \hat v^6}{4  \hat c_1^{18}}\right)+O( \hat c_2^7)\nn\\
\eea
\end{small}
Similarly for $a_{2,4,6}$ we get
\begin{small}
\bea
&&a_2( \hat v)=\frac{3  \hat c_2^2  \hat v}{8  \hat c_1^5}+ \hat c_2^3 \left(-\frac{15 M^2}{16  \hat c_1^6}
+\frac{105  \hat v^2}{16  \hat c_1^8}-\frac{25  \hat c_0  \hat v}{8  \hat c_1^7}\right)\nn\\
&&+ \hat c_2^4 \left(\frac{35  \hat v \left(7  \hat c_0^2-12 M^2\right)}{16  \hat c_1^9}
+\frac{175  \hat c_0 M^2}{16  \hat c_1^8}+\frac{1575  \hat v^3}{16  \hat c_1^{11}}
-\frac{1365  \hat c_0  \hat v^2}{16  \hat c_1^{10}}\right)\nn\\
&&+ \hat c_2^5 \left(\frac{1155  \hat v^2 \left(17  \hat c_0^2-15 M^2\right)}{32  \hat c_1^{12}}
-\frac{105  \hat c_0  \hat v \left(9  \hat c_0^2-65 M^2\right)}{16  \hat c_1^{11}}
-\frac{21 M^2 \left(105  \hat c_0^2-26 M^2\right)}{32  \hat c_1^{10}}
+\frac{45045  \hat v^4}{32  \hat c_1^{14}}-\frac{3465  \hat c_0  \hat v^3}{2  \hat c_1^{13}}\right)\nn\\
&&+ \hat c_2^6 \left(\frac{1155  \hat v \left(-204  \hat c_0^2 M^2+11  \hat c_0^4+41 M^4\right)}{64  \hat c_1^{13}}
+\frac{105105  \hat v^3 \left(5  \hat c_0^2-3 M^2\right)}{32  \hat c_1^{15}}
-\frac{15015  \hat c_0  \hat v^2 \left(7  \hat c_0^2-24 M^2\right)}{32  \hat c_1^{14}}\right.\nn\\
&&\left.+\frac{1155  \hat c_0 M^2 \left(18  \hat c_0^2-19 M^2\right)}{64  \hat c_1^{12}}
+\frac{315315  \hat v^5}{16  \hat c_1^{17}}-\frac{1996995  \hat c_0  \hat v^4}{64  \hat c_1^{16}}\right)
+O( \hat c_2^7)\nonumber\\
&&a_4( \hat v)=\frac{315  \hat c_2^4  \hat v}{128  \hat c_1^9}+ \hat c_2^5 \left(-\frac{2835 M^2}{256  \hat c_1^{10}}
+\frac{25641  \hat v^2}{256  \hat c_1^{12}}-\frac{5607  \hat c_0  \hat v}{128  \hat c_1^{11}}\right)\nonumber\\
&&+ \hat c_2^6 \left(\frac{231  \hat v \left(439  \hat c_0^2-666 M^2\right)}{256  \hat c_1^{13}}
+\frac{61677  \hat c_0 M^2}{256  \hat c_1^{12}}+\frac{693693  \hat v^3}{256  \hat c_1^{15}}
-\frac{573573  \hat c_0  \hat v^2}{256  \hat c_1^{14}}\right)+O( \hat c_2^7)\nn\\
&&a_6( \hat v)=\frac{51975  \hat c_2^6  \hat v}{1024  \hat c_1^{13}}+O( \hat c_2^7)
\eea
\end{small}
Inverting (\ref{a0H2}) we get
\bea
&&v_0(a)=a  \hat c_1+\frac{ \hat c_2 \left(a \left(2  \hat c_0-3 a\right)+M^2\right)}{2  \hat c_1}+
\frac{ \hat c_2^2 \left(3 a^2  \hat c_0-3 a^3-\frac{1}{2} a \left( \hat c_0^2-3 M^2\right)- \hat c_0 M^2\right)}{ \hat c_1^3}\nn\\
&&+\frac{ \hat c_2^3 \left(16  \hat c_0^2 \left(M^2-3 a^2\right)+ \hat c_0 \left(140 a^3-72 a M^2\right)
	+66 a^2 M^2-105 a^4+4 a  \hat c_0^3-M^4\right)}{8  \hat c_1^5}\nn\\
&&+\frac{ \hat c_2^4 }{8  \hat c_1^7}\left( \hat c_0 \left(-654 a^2 M^2+1005 a^4+17 M^4\right)
-32  \hat c_0^3 \left(M^2-3 a^2\right)\right.\nn\\
&&\left.+10  \hat c_0^2 \left(27 a M^2-52 a^3\right)+450 a^3 M^2-603 a^5-5 a  \hat c_0^4-24 a M^4\right)\nn\\
&&+\frac{ \hat c_2^5}{ \hat c_1^9}\left(-2  \hat c_0^2 \left(-228 a^2 M^2+345 a^4+7 M^4\right)
+\frac{3}{8}  \hat c_0 \left(-2060 a^3 M^2+2667 a^5+142 a M^4\right)\right.\nn\\
&&+8  \hat c_0^4 \left(M^2-3 a^2\right)+ \hat c_0^3 \left(\frac{805 a^3}{4}-105 a M^2\right)+\frac{7 a  \hat c_0^5}{8}\nn\\
&&\left.+\frac{3}{16} \left(-237 a^2 M^4+2295 a^4 M^2-2667 a^6+M^6\right)
\right)\nn\\
&&+\frac{ \hat c_2^6}{16  \hat c_1^{11}}\left(1040  \hat c_0^3 \left(-30 a^2 M^2+45 a^4+M^4\right)
-3  \hat c_0^2 \left(-30600 a^3 M^2+38906 a^5+2355 a M^4\right)\right.\nn\\
&&-8  \hat c_0 \left(-1815 a^2 M^4+15060 a^4 M^2-16947 a^6+14 M^6\right)
-256  \hat c_0^5 \left(M^2-3 a^2\right)\nn\\
&&\left.+105  \hat c_0^4 \left(45 a M^2-86 a^3\right)
-8970 a^3 M^4+56574 a^5 M^2-58104 a^7-21 a  \hat c_0^6+153 a M^6\right)+O( \hat c_2^7)\nn\\
\eea
Inserting these results into the general formulae (\ref{u_246_general}) for the corrections of $u$ we finally get
\begin{small}
\bea
&&v_2(a)=-\frac{3 a  \hat c_2^2}{8  \hat c_1^3}+\frac{ \hat c_2^3 \left(-39 a^2+19 a  \hat c_0+6 M^2\right)}{8  \hat c_1^5}\nn\\
&&+\frac{ \hat c_2^4 \left(798 a^2  \hat c_0-900 a^3-145 a  \hat c_0^2+261 a M^2-132  \hat c_0 M^2\right)}{16  \hat c_1^7}\nn\\
&&+\frac{ \hat c_2^5 \left(25230 a^3  \hat c_0-9024 a^2  \hat c_0^2+8388 a^2 M^2
	-20115 a^4-7752 a  \hat c_0 M^2+910 a  \hat c_0^3+1536  \hat c_0^2 M^2-141 M^4\right)}{32  \hat c_1^9}\nn\\
&&+\frac{ \hat c_2^6}{64  \hat c_1^{11}}\left(-312300 a^2  \hat c_0 M^2+713850 a^4  \hat c_0-378640 a^3  \hat c_0^2
+77760 a^2  \hat c_0^3+239400 a^3 M^2\right.\nn\\
&&\left.-444024 a^5+119610 a  \hat c_0^2 M^2-5145 a  \hat c_0^4
-11601 a M^4+6674  \hat c_0 M^4-13440  \hat c_0^3 M^2\right)+O( \hat c_2^7)\nonumber\\
&&v_4(a)=-\frac{297 a  \hat c_2^4}{128  \hat c_1^7}
+\frac{3  \hat c_2^5 \left(-3540 a^2+1577 a  \hat c_0+408 M^2\right)}{128  \hat c_1^9}\nn\\
&&+\frac{ \hat c_2^6 \left(415992 a^2  \hat c_0-494442 a^3-75287 a  \hat c_0^2+113265 a M^2
	-51600  \hat c_0 M^2\right)}{256  \hat c_1^{11}}+O\left( \hat c_2^7\right)\nonumber\\
&&v_6(a)=-\frac{50139 a  \hat c_2^6}{1024  \hat c_1^{11}}+O( \hat c_2^7)
\eea
\end{small}
  \end{appendix}
\providecommand{\href}[2]{#2}\begingroup\raggedright\endgroup

\end{document}